\documentclass[letterpaper,11pt]{JHEP3}

\usepackage{epsfig}

\def\sech{{\rm sech}}
\def\endignore{}
\def\ignore #1\endignore{} 

\def\ssJ{{\scriptscriptstyle J}}

\def\ssM{{\scriptscriptstyle M}}
\def\ssN{{\scriptscriptstyle N}}
\def\ssP{{\scriptscriptstyle P}}

\def\ssR{{\scriptscriptstyle R}}

\def\Box{{\hbox{$\sqcup$}\llap{\hbox{$\sqcap$}}}}

\def\be{\begin{equation}}
\def\ee{\end{equation}}
\def\bea{\begin{eqnarray}}
\def\eea{\end{eqnarray}}
\def\nn{\nonumber}

\def\exd{{\rm d}}
\def\pref#1{(\ref{#1})}
\def\endignore{}
\def\ignore #1\endignore{} 

\def\bd{\begin{displaymath}}
\def\ed{\end{diplaymath}}

\def\d{\mathrm{d}}

\def\cW{{\cal W}}
\def\cA{{\cal A}}

\def\cF{{\cal F}}

\def\hcW{{\hat{\cal W}}}

\def\hcA{{\hat{\cal A}}}

\def\ba{\begin{eqnarray}}
\def\ea{\end{eqnarray}}
\def\be{\begin{equation}}
\def\ee{\end{equation}}

\def\endignore{}
\def\ignore #1\endignore{} 
\def\Box{{\hbox{$\sqcup$}\llap{\hbox{$\sqcap$}}}}

\def\be{\begin{equation}}
\def\ee{\end{equation}}
\def\bea{\begin{eqnarray}}
\def\eea{\end{eqnarray}}
\def\nn{\nonumber}

\def\exd{{\rm d}}
\def\pref#1{(\ref{#1})}
\def\endignore{}
\def\ignore #1\endignore{} 

\def\bd{\begin{displaymath}}
\def\ed{\end{diplaymath}}

\def\d{\mathrm{d}}

\def\cW{{\cal W}}
\def\cA{{\cal A}}

\def\cF{{\cal F}}

\def\cC{{\cal C}}

\def\O{\mathcal{O}}

\def\nn{\nonumber}

\def\d{\mathrm{d}}

\def\({\left(}
\def\){\right)}

\def\nab{{\eta}_a}
\def\nbb{{\eta}_b}
\def\pab{{\omega}_a}

\def\la{\lambda_a}

\def\n{\eta}
\def\l{\lambda}

\def\hcW{{\hat{\cal W}}}

\def\hcA{{\hat{\cal A}}}

\def\cosh{{\rm cosh}}
\def\sinh{{\rm sinh}}
\def\tanh{{\rm tanh}}
\def\arcosh{{\rm arcosh}}
\def\arsinh{{\rm arsinh}}
\def\e{{e}}
\def\sign{{\rm sign}}

\def\d{ {\rm d} }

\def\bd{\boxdot}

\title{UV Caps and Modulus Stabilization
for 6D Gauged Chiral Supergravity}

\author{C.P. Burgess,${}^{1,2}$ D. Hoover${}^3$
and G. Tasinato${}^4$\\
${}^1$ Dept. of Physics \& Astronomy, McMaster University,
Hamilton ON, Canada, L8S 4M1. \\
${}^2$ Perimeter Institute for Theoretical Physics, Waterloo ON,
 Canada, N2L 2Y5. \\
${}^3$ Physics Department, McGill University, Montr\'eal QC,
Canada, H3A 2T8. \\
${}^4$ The Rudolph Peierls Centre for Theoretical Physics,
Oxford University, Oxford UK, OX1 3NP.\\
}

\date{}

\abstract {We describe an explicit UV regularization of the brane
singularities for all 4D flat configurations of 6D gauged chiral
supergravity compactified on axially symmetric internal spaces
(for which the general solutions are known). All such solutions
have two or fewer co-dimension two singularities, which we resolve
in terms of microscopic co-dimension one cylindrical 4-branes,
whose interiors are capped using the most general possible 4D flat
solution of the 6D field equations. By so doing we show that such
a cap is always possible for any given bulk geometry, and obtain
an explicit relationship between the properties of the capped
4-branes and the various parameters which describe the bulk
solution. We show how these branes generically stabilize the size
of the extra dimensions by breaking the scale invariance which
relates classical solutions to 6D supergravity, and we compute the
scalar potential for this modulus in the 4D effective theory. The
lifting of this marginal direction provides a natural realization
of the Goldberger-Wise stabilization mechanism in six dimensions.}

\begin{document}

\section{Introduction}

Six-dimensional supergravity \cite{SS,NS,6DSugra,HiDSugra} has
recently emerged as being a useful theoretical workshop within
which to investigate phenomena which often generalize to systems
having even more dimensions. Six dimensions are ideal for this
purpose inasmuch as there are enough dimensions to permit the
physics of most interest --- such as chiral fermions \cite{SS},
intricate Green-Schwarz anomaly cancellation
\cite{6DAnomalyCancellation} and flux-stabilized compactifications
\cite{SS,Susha}. Yet there are also few enough dimensions to allow
the relevant field equations to be solved explicitly, allowing a
detailed exploration of features which are more complicated to
investigate in a 10- or 11-dimensional context.

Interest in six dimensions has been further sharpened by the
recognition that it can provide significant insights into
phenomenological problems in its own right. Prominent among these
is the potential for having extra dimensions large enough to be
relevant to precision measurements of gravity on micron length
scales \cite{ADD}, and the potential of having the scale of
gravity be as low as the weak scale \cite{WeakScaleGravity}. Its
supersymmetric version, with supersymmetry broken by branes,
provides a realization of weak-scale supersymmetry breaking which
does {\it not} predict the existence of superpartners for standard
particles like the electron \cite{6DSUSYBreaking}, and so whose
implications for colliders differs considerably from standard
supersymmetric scenarios. It may yet provide an attractive
approach to the cosmological constant problem
\cite{SLED1}--\cite{SLEDpheno}, by building on the observation
that higher-dimensional theories can break the link on which the
cosmological constant problem rests: the link between the 4D
vacuum energy density (which we believe to be large) and the
curvature of 4D spacetime (which we observe to be small)
\cite{5DSelfTune} -- \cite{6DNonSUSYSelfTunex}.

The study of the physics of 6D supergravity was considerably
advanced by the discovery of the most general class of
compactifications to 4D flat space on an axially symmetric extra
dimensional geometry \cite{GGP,SLED2} which involve only the
fields of the supergravity multiplet itself. Because these are the
most general such solutions, they allow a more systematic study of
the circumstances under which the observed, noncompact four
dimensions are flat. In particular, these solutions are found to
be singular at one or two locations within the extra dimensions
\cite{GGPplus}, with the singularities being interpreted as
representing the back-reaction of codimension-two 3-branes whose
presence sources the fields described by the bulk fields under
consideration. Of pressing interest is the identification of the
kinds of brane properties which give rise to geometries with four
flat observed dimensions.

Unfortunately, the characterization of the required brane
properties is more complicated for codimension-two objects than it
is for the more familiar codimension-one configurations familiar
from Randall-Sundrum compactifications \cite{RS}. This is because
the bulk fields sourced by higher codimension objects generically
diverge at the positions of these objects. For this reason all
detailed connections between bulk and brane properties have so far
relied on the use of `thick' branes -- {\it i.e.} explicit models
of the internal brane structure which allow the bulk-field
singularities to be resolved, and smoothed out
\cite{Gian1,6DSmoothing,Papazoglou,OtherCap}.

Our purpose in the present paper is to systematize this smoothing
analysis to the general class of 4D flat solutions known for
axially-symmetric internal geometries. We do so in order to
provide a sufficiently general class of singularity resolutions to
allow a meaningful mapping to be made between the properties of
the resolved branes and those of the bulk geometries which they
source. We resolve the bulk-field singularities at the source
3-branes by cutting off the bulk geometry with an explicit (but
broad) class of cylindrical 4-branes which consistently couple to
all of the relevant bulk fields. Their interiors are then capped
off using the most general smooth, 4D-flat and cylindrically
symmetry solutions to the same 6D supergravity equations as are
solved by the bulk configurations.

Our main result is to provide explicit relations between the
properties of the 4-branes (and their capped geometries) and those
of the external bulk, a connection which pays at least two
dividends.
\begin{itemize}
\item First, by sharpening the general relations between the brane
and the bulk, our results provide the tools required to
definitively explore the sensitivity of bulk properties to the UV
structure on the source branes.

\item Second, because the capped branes generically break the
classical degeneracy between re-scaled bulk geometries, their
presence lifts this degeneracy and so provides a stabilization
mechanism which relates the size of the extra dimensions to
properties of the source branes. This stabilization mechanism can
be regarded as a particular form of the general Goldberger-Wise
mechanism \cite{GW} which arises particularly naturally within 6D
supergravity.
\end{itemize}

Our presentation of these results proceeds as follows. Next, in
\S2, we review the general 4D flat, cylindrically symmetric
solutions of ref.~\cite{GGP}, and use these to identify the form
taken by the smooth geometries which cap the interiors of the
cylindrical 4-branes. \S3 then follows with a detailed discussion
of the matching conditions which apply at the position of the
4-branes, and use these to identify the relationships which must
exist between the parameters of the bulk solutions and those which
govern the capped geometries and the intervening 4-branes. \S4
then focusses on the implications of these relations for the
parameters which govern the sizes of the bulk and capped
geometries, and identify the choices which must be made on the
branes in order to ensure a large hierarchy between the size of
the bulk and the size of the `thick' branes. Some conclusions are
summarized in \S5.

\section{Bulk solutions to 6D chiral supergravity}

We next review the properties of the field equations of 6D gauged
chiral supergravity \cite{NS,6DSugra,HiDSugra}, and present the
most general solutions to these equations for which the induced
geometry of the non-compact 4D directions is flat
\cite{GGP,SLED2,GGPplus}.

\subsection{6D field equations}

The action whose variation gives the field equations of interest
is  part of the Lagrangian density for 6D chiral gauged
supergravity, and is given by\footnote{The curvature conventions
used here are those of Weinberg's book \cite{GandC}, and differ
from those of MTW \cite{MTW} only by an overall sign in the
Riemann tensor.}
\be \label{6DSugraAction}
    \frac{{\cal L}}{\sqrt{-g}} = - \frac{1}{2 \kappa^2} \,
    g^{\ssM\ssN} \Bigl[ R_{\ssM\ssN} + \partial_\ssM \phi \,
    \partial_\ssN \phi \Bigr]
    - \frac14 \, e^{-\phi} \, F_{\ssM\ssN} F^{\ssM\ssN}
    - \frac{2g^2}{\kappa^4} \; e^\phi  \,,
\ee
where $\phi$ is the 6D scalar dilaton, and $F = \exd A$ is the
field strength for the gauge potential, $A_\ssM$, whose flux in
the extra dimensions is what stabilizes the compactifications. The
couplings $g$ and $\kappa$ have dimensions of inverse mass and
inverse mass-squared, respectively. (We keep $\kappa^2$ explicit
for ease of comparison with the various conventions which are used
in the literature.)

These expressions set some of the bosonic fields of 6D
supergravity to zero, as is consistent with the corresponding
field equations (see however ref.~\cite{HypersNonzero} for
solutions which do not make this assumption). The field equations
for $\phi$, $A_\ssM$ and $g_{\ssM\ssN}$ are:
\bea \label{fieldequations}
    && \Box \, \phi  +
    \frac{\kappa^2}{4} \, e^{-\phi} \; F_{\ssM\ssN} F^{\ssM\ssN}
    - \frac{2 \,g^{2}}{\kappa^2} \, e^{\phi} = 0\nn \\
    && D_\ssM \Bigl(e^{ - \phi} \, F^{\ssM\ssN} \Bigr) = 0 \\
    && R_{\ssM\ssN} + \partial_\ssM\phi \, \partial_\ssN\phi
    +  \kappa^2 e^{- \phi} \;
    F_{\ssM\ssP} {F_\ssN}^\ssP
    + \frac12 \, (\Box\,
    \phi)\, g_{\ssM\ssN} = 0 \,.\nn
\eea

The lagrangian density, eq.~\pref{6DSugraAction}, has an important
classical scaling property which plays a role in what follows: it
re-scales as ${\cal L} \to e^{2\omega} {\cal L}$ when the fields
undergo the constant re-scalings $g_{\ssM\ssN} \to e^\omega \,
g_{\ssM\ssN}$, $e^\phi \to e^{\phi - \omega}$ and $A_\ssM \to
A_\ssM$. Although it is not a symmetry of the action, it is a
symmetry of the field equations and so its action always relates
classical solutions to one another.

There is an ever-growing literature on the exact solutions to
these equations, describing static compactifications of 6D down to
4D \cite{NS,SLED1,SLED2,GGP,GGPplus}, as well as 4D de Sitter
solutions \cite{6DdSSUSY}, time-dependent solutions to the
linearized equations \cite{Linearized,KickRB} and exact scaling
solutions \cite{Scaling}. Our interest in what follows is in those
which are cylindrically symmetric and asymptotically flat.

\subsubsection*{Boundary contributions}

For later purposes we also record here the additional
Gibbons-Hawking term \cite{GibbonsHawking} with which the above
action must be supplemented when the field equations are
investigated in the presence of boundaries. If the 6D spacetime
region of interest, $M$, has a 5D boundary, $\Sigma = \partial M$,
then the full action for the bulk fields is
\be \label{GHTerm}
    S = \int_M \exd^6x \; {\cal L}
    - \int_\Sigma \sqrt{-\gamma} \,
    K \,,
\ee
where\footnote{In the following we use capital latin letters for
6D indices $(M,N)$ which run from $0\ldots5$; lower-case latin
letters for 5D indices $(m,n)$ which run over the 4-brane
directions, $0\ldots4$; and greek letters $(\mu,\nu)$ for 4D
indices which run over the noncompact dimensions, $0\ldots3$.}
$\gamma_{mn}$ denotes the induced metric on $\Sigma$ and $K =
\gamma^{mn} K_{mn}$, is the trace of the extrinsic curvature
tensor, $K_{mn}$, on $\Sigma$.

\subsection{General bulk solutions}

The most general axially-symmetric 4D-flat solutions to these bulk
equations of motion are given by metrics of the form
\be \label{eommetric}
    {\rm ds^2} = e^{\omega-p}\cW^2(\eta) \eta_{\mu \nu}
    \d x^{\mu} \exd x^{\nu}
    + \cA^2(\eta) \cW^8(\eta) \exd \eta^2
    + \cA^2(\eta) \exd
    \psi^2 ,
\ee
where $x^{\mu}$ label the four noncompact dimensions, and
$\{\eta,\psi\}$ are coordinates in the two extra dimensions,
satisfying the periodicity condition $0 \leq \psi \leq 2\pi$.
Solving the field equations, using for simplicity units for which
$\kappa^2 =1$, then gives\footnote{Beware that ref.~\cite{GGP}
instead uses $\kappa^2 = \frac12$.} the following formulae for the
unknown functions $\cA(\eta)$ and $\cW(\eta)$ \cite{GGP}

\bea \label{eom2}
    \cW^4 &=& \Big| \frac{q \l_2}{2g\l_1}\Big|
    \frac{ \cosh[\l_1 (\n-\xi_1)] }
    { \cosh[\l_2( \n - \xi_2 )]} \nn \\
    \cA^{-4} &=& \left| \frac{2 g \, q^3}{\l_1^3 \l_2}\right|
    \e^{-2(\l_3 \n +\omega)} \, \cosh^3[\l_1(\n-\xi_1)] \,
    \cosh[ \l_2 ( \n -\xi_2)] \nn \\
    \hbox{while} \quad
    \e^{-\phi} &=& \cW^2 \, e^{\l_3 \n + \omega}
    \quad \hbox{and} \quad
    F_{\n \psi} = \frac{q \cA^2}{\cW^2}
    \, \e^{-\l_3 \n-\omega} \,.
\eea
Here $q$, $\omega$, $\l_i$ ($i = 1,2,3$) and $\xi_a$ ($a = 1,2$)
are arbitrary integration constants, subject only to the
constraint $\l_2^2=\l_1^2+\l_3^2$. The role of the constant $p$ is
discussed further below. Notice that the signs of both $\l_1$ and
$\l_2$ are irrelevant in these solutions, and so without loss of
generality we take $\l_1>0$ and $\l_2>0$. Also, since in all
subsequent equations it is only the magnitude of $g$ which
appears, we simplify notation by writing $g$ instead of $|g|$.

For later convenience it is useful to display here the form of a
gauge potential, $A_\ssM$, whose differentiation gives the above
field strength, $F_{\eta\psi}$:
\be
    A_{\psi} = \frac{\lambda_1 }{q}\,\Bigl(
    \tanh\left[\lambda_1\left(\eta-\xi_{1}
    \right)\right] + \alpha\Bigr) \,,
\ee
where $\alpha$ is an arbitrary integration constant.

\subsubsection*{The parameters $p$, $\omega$ and $\xi_1$}

The parameters $p$ and $\omega$ appearing in eq.~\pref{eommetric},
may appear unfamiliar to aficionados of ref.~\cite{GGP}, since
they are not seen in the solutions given there. They do not do so
because each corresponds to a symmetry direction, and so for
simplicity they are both removed in ref.~\cite{GGP}. We reinstate
them here because we shall find that their removal is not
similarly possible for the bulk {\it and} for the cap geometries
which we consider shortly.

The symmetry corresponding to additive shifts of the variable
$\omega$ is just the classical scale invariance of the field
equations discussed above. The symmetry corresponding to $p$ is
similarly given by rigidly re-scaling the 4D metric, $g_{\mu \nu}
\to e^{-p} \, g_{\mu \nu}$. This can be seen to be a symmetry of the
field equations, eqs.~\pref{fieldequations}, once these are
restricted to the ansatz of eq.~\pref{eommetric} together with
$\phi = \phi(\eta)$ and $A_\psi = A_\psi(\eta)$. (Notice to this
end that this ansatz implies in particular that the 4D part of the
Ricci tensor, $R_{\mu\nu} = {R^\ssM}_{\mu \ssM \nu}$,
scales in the same way as does the 4D metric, $g_{\mu\nu}$.)

There is a third parameter in eqs.~\pref{eom2}, say $\xi_1$, which
could also have been eliminated in this way, since it can be
removed by a suitable choice of the origin for the coordinate
$\eta$. More formally, the field equations,
eqs.~\pref{fieldequations}, enjoy the symmetry $\eta \to \eta +
\delta$, for constant $\delta$, although the solutions,
eqs.~\pref{eom2}, do not. So applying such a shift to any given
solution generates a one-parameter set of new solutions. In fact,
inspection shows that the new solution obtained differs
from the original one simply by making the changes
\be \label{equivchanges}
    \xi_i \to \xi_i + \delta  \,, \qquad
    \omega \to \omega - \l_3 \, \delta
    \quad \hbox{and} \quad
    p \to p + \l_3 \, \delta \,.
\ee
This fact is important later since it tells us that one of the
parameters which governs the bulk solutions can be arbitrarily
removed by making an appropriate choice for the origin of
coordinates for $\eta$.
This means that one of these parameters, say $\xi_1$, has no
physical meaning and so one might wonder why we include it. The
reason is that when branes are included, it is useful to use the
$\eta$-shift symmetry to place them at convenient locations. Since
we have then used up this symmetry, the parameter $\xi_1$ takes on
a physical significance to do with the brane location.

\subsubsection*{Singularities}

The bulk solutions of eqs.~\pref{eom2} are regular for all finite
$\eta$, but generically are singular as $\eta \to \pm \infty$. The
nature of these singularities is most easily seen by transforming
to proper distance, $\exd \rho = \cA \, \cW^4 \, \exd \eta$. In
this limit the extra-dimensional part of the metric becomes $\exd
\rho^2 + C\rho^a \exd \psi^2$, which has a curvature singularity
at $\rho \to 0$ provided $a \ne 2$. If $a = 2$, the geometry has a
conical singularity when $a = 2$ and $C \ne 1$. When $a = 2$ and
$C = 1$ the solution is completely nonsingular at $\rho = 0$. (The
only solution having no singularities at all is the Salam-Sezgin
solution of ref.~\cite{SS}.)

Inspection of the asymptotic forms of eqs.~\pref{eom2} shows that
both of the singularities ({\it i.e.} those at $\eta \to \pm
\infty$) are conical if and only if $\lambda_1 = \lambda_2 \equiv
\lambda$ (and so $\lambda_3 = 0$). For the 4D flat solutions
considered here either both singularities are conical or neither
of them are (see ref.~\cite{6DdSSUSY} for non-flat solutions
having only one conical singularity). When $\xi_1 \ne \xi_2$ the
geometries with conical singularities are generically warped,
giving the solutions of ref.~\cite{SLED2}. However, if $\xi_1 =
\xi_2$ the conical solutions degenerate into the unwarped `rugby
ball' solutions of ref.~\cite{SLED1}.

Physically, the singularities at $\eta \to \pm \infty$ indicate
the presence of codimension-two source branes at these positions,
with the singular behaviour arising because of the back-reaction
of these branes onto the bulk fields. Furthermore, the precise
kind of singularity is expected to be related to the properties of
these source branes \cite{NavSant,GGPplus,Scaling}, with branes
that source the dilaton field $\phi$ typically giving rise to a
bulk scalar field configuration which diverges at the brane
position, and so whose energy density can give rise to curvature
singularities there.

Our goal in this section is to sharpen this connection, by
relating more precisely the integration constants of the bulk
solutions to the properties of the two source branes. We do so by
explicitly resolving the singularities at $\eta \to \pm \infty$ in
terms of a model of the microscopic structure of these two
codimension-two branes.

\subsection{Capped solutions}

To this end we model each of the source branes as a cylindrical
codimension-one 4-brane, situated at a fixed value of $\eta$,
whose interior is filled in with one of the above bulk solutions
that is nonsingular everywhere within the interior of the
cylinder.

Consider then pasting together the following two metrics, along
the 4+1 dimensional surface at $\eta = \eta_a$:
\bea
     \label{capsolns}
    {\rm d\hat{s}^2} &=& e^{\omega_a - p_a}
    \hcW^2(\eta) \eta_{\mu \nu}
    \d x^{\mu} \exd x^{\nu}
    + \hcA^2(\eta) \hcW^8(\eta) \exd \eta^2
    + \hcA^2(\eta) \exd
    \psi^2, \quad -\infty < \eta \leq \nab, \nn \\
    {\rm ds^2} &=& e^{\omega} \cW^2(\eta) \eta_{\mu \nu}
    \d x^{\mu} \exd x^{\nu}
    + \cA^2(\eta) \cW^8(\eta) \exd \eta^2
    + \cA^2(\eta) \exd
    \psi^2, \quad\qquad\; \nab \leq \eta \leq \nbb \nn,
\eea
with a similar splicing being performed at $\eta = \eta_b$ onto a
nonsingular cap geometry which is defined for $\eta_b < \eta <
\infty$. Codimension-one 4-branes will be located at the two
boundaries $\eta=\eta_a$ and $\eta = \eta_b$, whose properties we
determine below by using the appropriate jump conditions. Notice
that we use the freedom to re-scale coordinates to set $p = 0$ in
the bulk geometry (for $\eta_a < \eta < \eta_b$), but having done
so we cannot also remove the dimensionless parameter $p_a$ (or
$p_b$) in the cap region.

For convenience we make here the choice that the coordinate
location of the brane in the bulk coordinate system, $\eta_a$, is
the same as its location in the cap coordinate system,
$\hat{\eta}_a$. There is generically no reason for these two
numbers to be the same, but as discussed earlier we may use the
shift $\eta$-shift symmetry, eq.~\pref{equivchanges}, to enforce
$\eta_a = \hat{\eta}_a$. Having done this, we see that one of the
previously unphysical parameters in the cap, say $\xi_{1a}$, takes
on physical significance as it replaces $\hat{\eta}_a$.

For the cap solution which applies for $\eta < \eta_a$ we take one
of the geometries of eqs.~\pref{eom2}, subject to the condition
that it be singularity free as $\eta \to - \infty$. This is only
possible if it satisfies $\lambda_3 = 0$ --- and so $\lambda_1 =
\lambda_2 \equiv \lambda_a$ --- leading to the form
\bea
    \label{eom1}
    \e^{-\hat{\phi}} &=& \hcW^{2} \, e^{\pab}, \nn \\
    \hcW^4 &=& \Big| \frac{q_a }{2g_a
    }\Big| \frac{ \cosh[\l_a (\n-\xi_{1a})] }
    { \cosh[\l_a( \n - \xi_{2a} )]} \nn \\
    \hcA^{-4} &=& \Big| \frac{2 g_a
    \, q_a^3}{\l_a^4 }\Big| \e^{-2\pab } \,
    \cosh^3[\l_a(\n-\xi_{1a})]\,
    \cosh[ \l_a ( \n -\xi_{2a} )] \nn \\
    \hat{F}_{\eta \psi} &=&
    \frac{q_a \hcA^2}{\hcW^2} \, \e^{-\pab} \,.
\eea
Similarly to the bulk case, we are free to take $\la
> 0$. Also, as was done with $g$, for simplicity we write $g_a$
in place of $|g_a|$. We are led in this way to the following 7
integration constants describing each capped geometry: $\la$,
$p_a$, $q_a$, $\omega_a$, $\xi_{1a}$, $\xi_{2a}$ and $\eta_a$. By
contrast, the constant $g_a$ is {\it not} an integration constant,
but is the $U_\ssR(1)$ gauge coupling which appears in the bulk
action whose equations of motion govern the solutions of interest.
Although we keep $g_a$ and $g$ distinct in what follows, this is
not crucial for our results, and one could instead choose to use
the same action for the cap regions and the bulk between the two
branes: $g_a = g$.

Requiring the cap geometry to be smooth for $\eta\to-\infty$
imposes the following relation amongst the integration constants:
\be \label{constfcap}
    |q_a| = 2 \la g_a
    \,e^{\la\left(\xi_{2a} -\xi_{1a} \right)} .
\ee
In what follows we regard this last equation as fixing the
combination $\xi_{2a} - \xi_{1a}$. When the result satisfies
$\xi_{1a} \ne \xi_{2a}$ the capped geometry is warped, and we
refer to it as a `tear drop'. In the special case $\xi_{1a} =
\xi_{2a} \equiv \xi_a$ --- {\it i.e.} when $|q_a| = 2 \la g_a$ ---
the cap geometry instead degenerates into a hemisphere.

\subsubsection*{Parameter counting}

For future convenience it is useful at this point to count the
number of integration constants associated with each of the
solutions.

\begin{itemize}
\item {\it The Bulk:} Using the coordinate freedom to re-scale
$g_{\mu \nu}$ and to shift $\eta$, we may set $p = 0$ and fix
$\xi_2$ to a particular value. This leaves the general bulk
solutions characterized by the 5 integration constants $\l_1$,
$\l_2$, $\xi_2$, $q$ and $\omega$.
\item {\it The Caps:} The same coordinate freedom cannot again be
used to similarly simplify the teardrop cap geometries for the
regions $\eta < \eta_a$ and $\eta > \eta_b$. Once one parameter
({\it e.g.} $\xi_{2a}$) is used to ensure the cap geometry is
everywhere smooth -- {\it c.f.} eq.~\pref{constfcap} -- each cap
is therefore described by 6 parameters. For the cap at $\eta <
\eta_a$ these are $\l_a$, $\xi_{1a}$, $q_a$, $p_a$ and $\omega_a$,
together with the 4-brane location, $\nab$. For the cap at $\eta >
\eta_b$ we instead have $\l_b$, $\xi_{1b}$, $q_b$, $p_b$,
$\omega_b$ and $\eta_b$.
\end{itemize}
To these parameters we must also add those that characterize the
4-brane action, as is discussed in some detail in the next
section.

We do not include the gauge potential integration constant,
$\alpha$, in the above counting because we handle its matching
conditions separately in what follows. Besides $\alpha$, the gauge
potential also potentially hides other moduli describing how the
background gauge field is embedded within the full gauge group.
This can show up in the present analysis by making the gauge
coupling constant, $e$, associated with the background gauge field
potentially different from the coupling $g$ which appears in the
supergravity action, eq.~\pref{6DSugraAction}, and so also in the
solutions, eqs.~\pref{eom2} \cite{SLED1,SLED2}.

\section{Matching conditions}

We next impose the matching conditions which apply across the
4-brane position, where the cap geometry meets that of the bulk.
These come in two types: continuity of the fields $g_{\ssM\ssN}$,
$A_\ssM$ and $\phi$ across $\eta = \eta_a$, and jump conditions
which relate the discontinuity in the derivatives of these fields
to properties of the 4-brane action.

\subsection{Continuity conditions}

Continuity of the bulk fields at each brane position provides 4
conditions among the parameters which define the caps. For
instance, continuity across the 4-brane situated at $\eta_a$
gives:
\be \label{J1}
    e^{\omega_a - p_a} \hcW^2(\nab) =
    e^{\omega} \cW^2(\nab) \,, \quad
    \hcA^2(\nab) = \cA^2(\nab) \,, \quad
    \hat\phi(\nab) = \phi(\nab) \,
\ee
and
\be \label{J1gauge}
    \hat A_\psi(\nab) = A_\psi(\nab) \,.
\ee

After some simplification, the three conditions of eqs.~\pref{J1}
reduce to the following relations amongst the parameters of the
capped and bulk solutions
\begin{eqnarray}
    \frac{\cosh [{\lambda_1\left(\eta_a - \xi_1 \right)}]}{
    \cosh[{\lambda_a \left(\eta_a - \xi_{1a} \right)}]}&=&
    \Big| \frac{\lambda_1 q_a}{\lambda_a q} \Big| \label{continuity1}\\
    \frac{\cosh[{\lambda_2 \left(\eta_a - \xi_2\right)}]}{\cosh[{
    \lambda_a \left(\eta_a - \xi_{2a} \right)}]}
    &=& \Big| \frac{g_a \,\lambda_2}{g\,\lambda_a} \Big|
    \, e^{2(\omega-\omega_a + \lambda_3 \eta_a) }\label{continuity2}\\
    p_a&=& \lambda_3 \eta_a \,, \label{continuity3}
\end{eqnarray}
with a similar set of relations holding for brane $b$. As we see
below in more detail in subsection (\ref{solvconcon}), these
equations can be regarded as fixing the three parameters $p_a$,
$\xi_{2a}$ and $q_a$, leaving $\lambda_{a}$, $\omega_a$ and
$\eta_a$ free.

\subsubsection*{Topological constraint}

We treat the continuity condition for the gauge potential
separately, because of a topological subtlety which arises in this
case. Recall that the gauge potential for the bulk and capped
regions can be written in the form
\begin{eqnarray}
    A_{\psi}&=&\frac{\lambda_1 }{q}\,\Bigl(
    \tanh\left[\lambda_1\left(\eta-\xi_{1}
    \right)\right]+\alpha\Bigr)
    \hskip 1cm \eta_a< \eta< \eta_b \nn\\
    \hat A_{\psi} &=& \frac{\lambda_a}{q_a}\,\Bigl(
    \tanh\left[\lambda_a \left(\eta-\xi_{1a}
    \right)\right] + 1\Bigr)
    \hskip 1cm -\infty < \eta < \eta_a
\end{eqnarray}
where the integration constant is chosen in the capped region to
ensure that $A_\psi$ vanishes as $\eta \to - \infty$, as is
required for a nonsingular gauge potential. The same reasoning
applied to the second capped region similarly gives
\begin{eqnarray}
    A_{\psi}&=&\frac{\lambda_1 }{q}\,\Bigl(
    \tanh\left[\lambda_1\left(\eta-\xi_{1}
    \right)\right]+\alpha'\Bigr)
    \hskip 1cm \eta_a< \eta< \eta_b \nn\\
    \hat A_{\psi} &=& \frac{\lambda_b}{q_b}\,\Bigl(
    \tanh\left[\lambda_b \left(\eta-\xi_{1b}
    \right)\right] - 1\Bigr)
    \hskip 1cm \eta_b < \eta < \infty
\end{eqnarray}
where the integration constant is in this case chosen in the
capped region to ensure that $A_\psi$ vanishes as $\eta \to +
\infty$.

Naively we would determine $\alpha$ and $\alpha'$ by working
within a gauge for which $A_\psi$ is continuous for all $\eta$.
However, the crucial point is that there is in general a
topological obstruction to making such a choice for $A_\ssM$
everywhere. Instead we choose a gauge for which $A_\psi(\eta_a) =
\hat A_\psi(\eta_a)$ and $A_\psi(\eta_b) = \hat A_\psi(\eta_b)$,
and use these conditions to determine $\alpha$ and $\alpha'$. But
then $\alpha'$ and $\alpha$ cannot be taken to be equal on the
region of overlap, $\eta_a < \eta < \eta_b$, but must differ
instead by a gauge transformation. Following standard arguments,
this leads to the quantization condition
\be \label{quantizn}
    \frac{\lambda_1}{q}\, \left(\alpha - \alpha'\right)
    = \frac{N}{e}
\ee
where $N$ is an integer, and $e$ is the gauge coupling for the
background gauge field (which need not equal $g$ if the background
flux is not the one gauging the specific $U_\ssR(1)$ symmetry).

We find in this way that eq.~\pref{quantizn} implies the following
quantization condition on the various parameters:
\bea \label{genqua}
    \frac{N}{e} &=& \frac{\l_1}{q} \Bigl(
    \tanh[\l_1(\nbb-\xi_1)]
    - \tanh[\l_1(\nab-\xi_1)]
    \Bigr) \nn\\
    && + \frac{\l_a}{q_a} \Bigl( \tanh[\l_a
    (\nab-\xi_{1a})] + 1 \Bigr)
    - \frac{\l_b}{q_b} \Bigl( \tanh[\l_b(
    \nbb-\xi_{1b})]-1 \Bigr).
\eea
This generalizes to the case of thick branes the well-known Dirac
quantization condition $N/e = 2 \l_1/q$ \cite{SLED1,GP}, which is
retrieved from eq.~\pref{genqua} in the thin-brane limit obtained
by taking $\nab\to -\infty$ and $\nbb\to +\infty$.

Such arguments show that in general the continuity of the gauge
potential across the two 4-branes, $\eta = \eta_a$ and $\eta =
\eta_b$, determines the integration constants, $\alpha$ and
$\alpha'$ which are specific to the gauge potentials. But the
topological constraint then implies a single additional condition,
eq.~\pref{genqua}, which relates the bulk parameters, $\l_1$,
$\xi_1$ and $q$, to the undetermined brane quantities, $\nab$,
$\xi_{1a}$, $\nbb$, $\xi_{1b}$ and the flux integer $N$.

\subsection{Jump conditions}

Having examined the continuity conditions, we next examine the
relevant jump conditions which govern the discontinuity of
derivatives of the bulk fields across the brane positions at $\eta
= \eta_a$ and $\eta = \eta_b$. These junction conditions relate
any such a discontinuity to the dependence of the intervening
4-brane action, ${\mathcal S}$, on these bulk fields, and may be
derived by integrating the equations of motion across a narrow
interval around the 4-brane position: $\eta_a - \epsilon < \eta <
\eta_a + \epsilon$, with $\epsilon$ taken negligibly small.
Specialized to the metric these conditions are known as the Israel
junction conditions \cite{Israel}.

One finds in this way
\be \label{genjc}
    \lbrack {\mathbb K}_{mn} \rbrack_\ssJ =
    -T_{mn} \,, \qquad
    \lbrack \sqrt{-g}\; e^{-\phi}
    F^{\eta m} \rbrack_\ssJ
    = -\frac{\delta {\mathcal S}}{\delta A_m} \quad
    \hbox{and}\quad
    \lbrack \sqrt{-g} \; \partial^{\eta}
    \phi \rbrack_\ssJ = - \frac{\delta
    {\mathcal S}}{\delta \phi} \,,
\ee
where we use the definition $\lbrack f(\eta) \rbrack_{\nab} \equiv
f(\nab+\epsilon)-f(\nab-\epsilon)$. Here we define $K =
\gamma^{mn} K_{mn}$ and ${\mathbb K}_{mn} = K_{mn} - \gamma_{mn}
K$, where $K_{mn}$ is the extrinsic curvature of the appropriate
4-brane surface.

\subsubsection*{4-Brane action}

In order to proceed we require an ansatz for the 4-brane action.
Consider therefore the following general choice
\be
    \label{4brane}
    {\mathcal S} = -\int_{\Sigma}
    \exd^5x \sqrt{-\gamma} \left[ V(\phi)
    + \frac{1}{2} \, U(\phi)
    ( {D}_m \sigma {D}^m \sigma)
    \right],
\ee
where $\gamma_{mn}$ is the induced metric on the brane, and
$V(\phi)$ and $U(\phi)$ are functions which determine the 4-brane
couplings to the 6D dilaton.

Following ref.~\cite{Gian1} we introduce a Stueckelberg field,
$\sigma$, living on the brane, whose gauge covariant derivative is
$D_m \sigma = \partial_m \sigma - e A_m$. We imagine this to be
the low energy effective action obtained by integrating out the
massive mode of some brane-localized Higgs field, $H = {v \ e}^{i
\sigma}$, where $v$ is an appropriate expectation value.
Physically, this field describes supercurrents whose circulation
can support changes in the background flux across the position of
the 4-brane.(We return to the necessity for including such a field
in subsequent sections.) The equation of motion for $\sigma$,
together with the periodicity requirement $\psi \simeq \psi +
2\pi$, allows us to write the background configuration for
$\sigma$ as
\be
    \sigma = k \, \psi,
\ee
for some integer $k \in {\mathbb Z}$.

With these choices the jump conditions, eqs.~\pref{genjc}, become
\bea\label{genjcmunu}
    \lbrack {\mathbb K}_{\mu\nu} \rbrack_\ssJ &=&
    -T_{\mu  \nu } \\ \label{genjcpsi}
    \lbrack {\mathbb K}_{\psi \psi} \rbrack_\ssJ  &=&
    -T_{\psi \psi} \\ \label{genjcAm}
    \lbrack \sqrt{-g}\; e^{-\phi}
    F^{\eta \psi} \rbrack_\ssJ
    &=& -e \, U \sqrt{-\gamma} \;{D}^{\psi}
    \sigma  \\ \label{genjcdilaton}
    \lbrack \sqrt{-g} \; \partial^{\eta}
    \phi \rbrack_\ssJ &=& \sqrt{-\gamma}
    \left[ \frac{dV}{d\phi} + \frac{1}{2} \,
    ({D}_m \sigma {D}^m
    \sigma) \frac{dU}{d\phi} \right] , 
\eea
where the energy-momentum tensor derived from the above action is
\bea
    T_{\mu \nu} &=& - e^{\omega} \left( \frac{\cW}{\cA} \right)^2
    \left[ \cA^2 V +
    \frac{1}{2} \, U (k - e A_{\psi} )^2 \right]
    \eta_{\mu \nu} \nn \\
    T_{\psi \psi} &=& - \left[ \cA^2 V
    - \frac{1}{2} \, U
    (k - e A_{\psi})^2 \right].
\eea
Here we see one reason for including the Stueckelberg field:
without the function $U$ the expressions for $T_{\mu\nu}$ and
$T_{\psi\psi}$ are not independent since their ratio would be
independent of parameters from the 4-brane action, leading to too
restrictive a set of geometries which could be described in the
bulk.

\subsubsection*{Evaluating the Junction Conditions}

We next specialize the junction conditions to the explicit bulk
fields discussed above. We first require the extrinsic curvature,
$K_{mn}$, evaluated on both sides of the brane. In the bulk
region, the unit normal to surfaces of constant $\n$ is
\be
    n_\ssM = \cA \cW^4 \, \delta^{\n}_\ssM
\ee
and so the extrinsic curvature is given by $K_{mn} = \nabla_m n_n
= -\cA \cW^4 \, \Gamma^{\eta}_{mn}$, where $\Gamma^{\n}_{mn}$ is
the Christoffel symbol calculated from the full 6D metric. We find
\bea
    {\mathbb K}_{\mu \nu} &=& - \frac{e^{\, \omega}}{\cA \cW^2}
    \left[ \frac{3\cW'}{\cW} +
    \frac{\cA'}{\cA} \right] \eta_{\mu \nu} \nn \\
    {\mathbb K}_{\psi \psi} &=& -\frac{4\cA \cW'}{\cW^5},
\eea
where primes denote differentiation with respect to $\eta$.
Similarly, in the cap regions we have
\bea
    \hat{{\mathbb K}}_{\mu \nu} &=&
    -\frac{e^{\, \omega_a-p_a}}{\hcA \hcW^2} \left[
    \frac{3\hcW'}{\hcW} + \frac{\hcA'}{\hcA} \right]
    \eta_{\mu \nu} \nn \\
    \hat{{\mathbb K}}_{\psi \psi} &=&
    -\frac{4\hcA \hcW'}{\hcW^5}.
\eea

\bigskip

\noindent $\bullet$ Evaluating the $(\mu \nu)$ Israel junction
condition at $\eta = \nab$ then gives\footnote{It is understood in
what follows that all functions depending on $\eta$ are evaluated
at $\eta = \nab$.}
\bea
    \label{Isr1}
    &&\left( \frac{\l_3}{2} + e^{-2(\omega-\omega_a
    + \l_3 \eta_a)} \la \, \tanh[\l_a(\nab-\xi_{2a} )]
    - \l_2 \, \tanh[\l_2(\nab-\xi_2)] \right) \nn \\
    && \hspace{150pt}
    = - \cW^4 \left[ (\cA V_a) + \frac{1}{2} \,
    \left( \frac{U_a}{\cA} \right) (k_a-e A_{\psi})^2 \right]
\eea
where the subscript `$a$' on $V$, $U$ and $k$ denotes the
corresponding 4-brane property specialized to the brane at $\eta =
\nab$.

\bigskip

\noindent$\bullet$ The $(\psi \psi)$ Israel junction condition
similarly gives
\bea
    \label{Isr2}
    &&  \left[ \l_1 \,
    \tanh[\l_1 (\nab-\xi_1)] -
    \l_2 \, \tanh[ \l_2 (\nab-\xi_2)] -
    e^{-2(\omega-\omega_a+\l_3 \eta_a)} \Big(
    \l_a \, \tanh[\l_a (\nab-\xi_{1a})] \right. \nn\\
    &&\qquad\qquad
    \left. - \l_a \, \tanh[ \l_a (\nab-\xi_{2a})] \Big)
    \right]  = - \cW^4 \left[ (\cA V_a) -
    \frac{1}{2} \, \left(\frac{U_a}{\cA}\right)
    (k_a-e A_{\psi})^2 \right] .
\eea
Taking the sum and the difference of these last two conditions
allows the isolation of conditions for $V_a$ and $U_a$ separately.
It is also easy to see that the resulting equations always admit
real solutions for any value of the bulk parameters and the brane
position.

\bigskip

\noindent $\bullet$ The junction condition for the gauge field
similarly evaluates to
\be \label{GJ}
    q- q_a e^{-2(\omega-\omega_a + \l_3 \eta_a)} =
    - e \, \cW^4 \left(\frac{U_a}{\cA}\right) (k_a - e A_{\psi}).
\ee

\bigskip

Notice that we can eliminate the two brane quantities, $U_a$ and
$V_a$, from the previous three jump conditions to obtain a
constraint that does not depend on 4-brane parameters. Indeed, by
subtracting eq.~\pref{Isr1} from eq.~\pref{Isr2}, and then
dividing the result by eq.~\pref{GJ}, we obtain the expression
\be
    \frac{\frac12 \lambda_3 +\lambda_a
    \tanh[\lambda_a\left(\eta_a-\xi_{1a}
    \right)]-\lambda_1
    \tanh[\lambda_1\left(\eta_a
    -\xi_1\right)]}{e^{-2(\omega-\omega_a
    + \l_3 \eta_a)}\,q_a-q}
    = -\frac{k_a}{e}+
    \frac{\l_a}{q_a}
    \Bigl( \tanh[\l_a(\nab-\xi_{1a})] + 1
    \Bigr) \,.\label{l_constraint}
\ee
An identical argument for brane $b$ similarly gives:
\be
    \frac{\frac12 \lambda_3 +\lambda_b
    \tanh[\lambda_b\left(\eta_b-\xi_{1b}
    \right)]-\lambda_1
    \tanh[\lambda_1\left(\eta_b
    -\xi_1\right)] }{e^{-2(\omega-\omega_b
    + \l_3 \eta_b)}\,q_b-q}
    = -\frac{k_b}{e}+
    \frac{\l_b}{q_b}
    \Bigl( \tanh[\l_b(\nbb-\xi_{1b})]
    - 1
    \Bigr) . \label{l_constraintb}
\ee

\bigskip

\noindent $\bullet$ By contrast, the dilaton junction condition
gives a condition on the $\phi$-derivatives of $U_a$ and $V_a$:
\bea \label{dilatonJ}
     && 2\l_3 + \l_1 \tanh[\l_1(\eta_a-\xi_1)]
     - \l_2 \tanh[\l_2(\eta_a -
     \xi_2)] - e^{-2(\omega-\omega_a+\l_3 \eta_a)} \Big( \la
     \tanh[\la(\nab-\xi_{1a})] \nn \\
     &&\qquad \qquad
     - \la \tanh[\la(\nab-\xi_{2a})] \Big)
     = -2\cW^4 \Big[ \cA \frac{\exd V_a}{\exd
   \phi} + \frac{1}{2\cA}
   \frac{\exd U_a}{\exd \phi} (k_a - e A_{\psi})^2
   \Big] \,,
\eea
which, using the $(\psi \psi)$ Israel jump condition, simplifies
to
\be \label{dilatonJ2}
   2\l_3 =  \cW^4  \Big[ \cA \left( V_a
   - 2 \frac{\exd V_a}{\exd \phi}  \right)
   - \frac{1}{2\cA}  \left(  U_a
   + 2 \frac{\exd U_a}{\exd \phi} \right)
   (k_a - e A_{\psi})^2
   \Big] \,.
\ee

\bigskip

\subsubsection*{Conditions for scale invariance}

Before proceeding it is useful to pause at this point to record
the unique choice for the functions $V_a$ and $U_a$ which
preserves the classical scaling symmetry of the bulk equations of
motion, corresponding to the transformation $\omega \rightarrow
\omega + \Delta$ and $\omega_a \rightarrow \omega_a + \Delta$.

Inspection shows that the continuity equations remain unchanged by
this transformation because $\omega$ and $\omega_a$ only appear
there in the combination $\omega-\omega_a$. The left-hand-sides of
the various jump conditions remain similarly unchanged. On the
right-hand-side, however, we see that $\cA$ transforms, and so
invariance requires $V_a(\phi)$ and $U_a(\phi)$ to transform in a
way which cancels the transformation of $\cA$. Such an invariant
choice for $U_a$ and $V_a$ is possible for the Israel and Maxwell
jump conditions, eqs.~\pref{Isr1}, \pref{Isr2}, and \pref{GJ},
because within these $U_a$ and $V_a$ only appear with $\cA$ in the
combinations $\cA V_a$ and $U_a/\cA$. It follows that preservation
of the classical scaling symmetry requires
\be \label{scaleinvUV}
    V_a = v_a e^{\phi/2}
    \qquad {\rm and} \qquad
    U_a = u_a e^{-\phi/2} \,,
\ee
in agreement with the analysis of ref.~\cite{SLED2}. Any other
choices for these functions necessarily breaks the classical scale
invariance of the problem.

It then remains to determine what invariance requires for the
dilaton jump condition, eq.~\pref{dilatonJ2}. When this is
specialized to the scale invariant case, eqs.~\pref{scaleinvUV},
the right-hand side degenerates to zero, giving the simple
condition $\l_3 = 0$. Besides imposing no new conditions on $U_a$
and $V_a$, this tells us that scale-invariant brane configurations
can only source bulk geometries satisfying $\l_3 = 0$, and hence
only having conical singularities. Since all of the geometries
having two conical singularities are 4D flat \cite{6DdSSUSY}, we
see in detail how the jump conditions enforce the connection
between scale invariance and 4D flatness.

\section{Applications}

Given the general bulk and cap solutions, and a complete set of
matching conditions, we may now see what the solutions to these
conditions tell us about bulk-brane dynamics in six dimensions. In
this section we use the above formalism to address two questions.
First: given a bulk geometry what kinds of caps are possible?
Second: given specific brane properties, what kinds of bulk are
generated? In particular, in this second case we ask how the
breaking of scale invariance by the branes can lead to the
stabilization of the extra-dimensional size.

Answering this last question allows us also to address an issue of
potential importance for phenomenology: what conditions must the
cap and bulk parameters satisfy in order to have a large hierarchy
between the volumes of the caps and the volume of the bulk? This
point is important when the regularizing 4-branes and caps are
regarded as specifying the microscopic structure of 3-branes that
sit at the singular points of the geometry.

\subsection{Capping a given bulk}\label{cappingbulk}

We begin by studying what kinds of caps can be used to smooth a
generic bulk solution. In this section we therefore regard the 5
bulk parameters $\lambda_1$, $\lambda_2$, $\xi_1$, $q$ and
$\omega$ as given (we remove both $p$ and $\xi_2$ using
appropriate coordinate conditions), and look for solutions for the
kinds of branes which can smooth the singularities at $\eta = \pm
\infty$.

We emphasize that our purpose here is simply to show that a
regularization procedure exists for any choice of bulk solution,
through an appropriate choice for the 4-branes and caps. We return
in subsequent sections to the relations which must exist between
the parameters governing the branes and caps, due to the
interpolation between them of a 4D flat bulk.

\subsubsection*{Parameter counting}

It is instructive to count parameters and constraints, to get a
sense of whether or not the problem of capping a given bulk is
over-determined. To this end it is worth distinguishing between
those parameters which are integration constants in the capped
region, and those which arise within the action, ${\cal S}$,
governing the 4-brane. We start by counting only those relations
which are independent of the 4-brane action, before returning to
those which are not.

\bigskip\noindent{\it ${\cal S}$-independent conditions:}
We have seen that each cap naively involves 7 integration
constants, $\lambda_a$, $\xi_{1a}$, $\xi_{2a}$, $p_a$, $q_a$,
$\omega_a$ and $\eta_a$ that are related by the smooth-geometry
condition, \pref{constfcap}, at each cap. Counting the two caps
this gives a total of $6+6=12$ independent cap integration
constants.

At each cap these parameters are subject to 3 continuity
conditions, eqs.~\pref{continuity1} -- \pref{continuity3}, as well
as the 1 jump condition, \pref{l_constraint} or
\pref{l_constraintb}, constructed by eliminating $U(\phi)$ and
$V(\phi)$ from eqs.~\pref{Isr1} -- \pref{GJ}. The topological
constraint then imposes one more overall relation which relates
the properties of the bulk to those of both caps, giving a grand
total of $4+4+1=9$ conditions. Barring other obstructions we then
expect to find a $12-9=3$-parameter family of capped geometries
which can match properly to the given bulk.

\bigskip\noindent{\it ${\cal S}$-dependent conditions:}
In addition to these are the parameters $U_a(\phi)$ and
$V_a(\phi)$ governing the 4-brane action, ${\cal S}$. For each
brane these two functions are related by the three remaining
conditions, eqs.~\pref{Isr1}, \pref{Isr2} and \pref{dilatonJ2}.
Solving the two linear equations, \pref{Isr1} and \pref{Isr2},
immediately gives $U_a$ and $V_a$ as explicit functions of
$\eta_a$: $U_a = U_a(\eta_a)$ and $V_a = V_a(\eta_a)$ (where we
suppress the dependence on the other cap and bulk parameters).

We are then left with one remaining relation: the dilaton jump
condition, eq.~\pref{dilatonJ2}. Since this requires knowing the
derivatives, $\exd U_a/\exd\phi$ and $\exd V_a/\exd\phi$, further
progress requires making some choices for the functional form of
these quantities.
\begin{itemize}
\item If $U_a$ and $V_a$ are both constant, then both are fixed by
eqs.~\pref{Isr1} and \pref{Isr2}. In this case the dilaton jump
condition, eq.~\pref{dilatonJ2}, imposes an additional 2
constraints (one at each cap) on the 3 cap integration constants
which remain to this point undetermined. We are then led to expect
a 1-parameter family of capped solutions.
\item If $U_a$ and $V_a$ preserve scale invariance, then $U_a =
u_a \, e^{-\phi/2}$ and $V_a = v_a \, e^{\phi/2}$, have 2 free
parameters. In this case the counting naively goes through as
above, with one change: although $u_a$ and $v_a$ are fixed by
solving the Israel junction conditions, eqs.~\pref{Isr1} and
\pref{Isr2}, the dilaton jump condition, eq.~\pref{dilatonJ2},
degenerates to $\lambda_3 = 0$ and so does not further constrain
any 4-brane or cap parameters. (None of these matching conditions
fix the scale symmetry $\omega \to \omega + \Delta$, $\omega_a \to
\omega_a + \Delta$ and $\omega_b \to \omega_b + \Delta$. However,
because we here regard the bulk parameter $\omega$ to have been
specified this symmetry does not preclude the determination of
$\omega_a$ and $\omega_b$ in terms of $\omega$.) We are therefore
led in this case to 3 free parameters in the capped solution.
\item More general choices for $U_a$ and $V_a$ potentially involve
more parameters, and so allow more freedom of choice for the
capped geometry. For instance, if $U_a = u_a \, e^{s_a \phi}$ and
$V_a = v_a \, e^{t_a \phi}$, then the three conditions,
\pref{Isr1}, \pref{Isr2} and \pref{dilatonJ}, provide three
relations amongst the four parameters $u_a$, $v_a$, $s_a$ and
$t_a$, and in particular \pref{dilatonJ} no longer constrains the
parameters of the caps. In this case we'd expect a total of 5 free
parameters to describe the capped geometry.
\end{itemize}

Considerations such as these lead us to expect that capped
solutions of the type we entertain should exist for any given kind
of bulk geometry, barring an obstruction to solving the relevant
equations. Furthermore, we expect to find at least a 1-parameter
family of such solutions, and this has a simple physical
interpretation: in the absence of the topological constraint the
caps have 2 free parameters, corresponding to the freedom to
choose the positions, $\eta_a$ and $\eta_b$, where we choose to
position the two caps. The topological constraint can then impose
one relation amongst these two positions, relating them to the
quantum number, $N$, which governs the total amount of Maxwell
flux.

Notice that our counting here regards $U$ and $V$ as parameters to
be adjusted even though these arise within the brane action rather
than as integration constants in the solutions to the field
equations. So the existence of the caps requires these parameters
in the action to be tuned relative to one in a way which depends
on the properties of the given bulk solution. We also do not
distinguish here whether the solutions found give positive values
for $U$ and $V$, as would normally be required by positivity of
the kinetic energy associated with brane motion ($V$) and the
Stueckelberg field ($U$).

\subsubsection*{Freely-floating 4-branes}

The previous section takes the point of view that the
$\phi$-dependence of the 4-brane action can be arbitrarily
parameterized, with the parameters required to cap the given bulk
geometry being fixed in terms of the positions of the caps and
other variables. Another point of view is to ask for a 4-brane
action to be defined so that the same 4-brane action can be used
at {\it any} 4-brane position, for a given bulk geometry. As we
shall see, consistency also requires the cap geometry to be varied
as a function of the brane position. This approach is similar in
spirit to what is done for the actions of end-of-the-world branes
which mark the boundary of bulk spaces in discussions of the
AdS/CFT correspondence \cite{EoWRG}.

This amounts to asking that the $\eta_a$-dependence inferred by
solving eqs.~\pref{Isr1} and \pref{Isr2} for $U_a(\eta_a)$ and
$V_a(\eta_a)$ is completely given by the implicit
$\eta_a$-dependence which $U_a$ and $V_a$ inherit as functions of
$\phi(\eta_a)$ (with $\eta_a$-independent constants). That is, we
demand $U_a(\eta_a) = U_a[\phi(\eta_a)]$ and $V_a(\eta_a) =
V_a[\phi(\eta_a)]$. We call such a 4-brane action the `floating'
action which is defined by the given bulk and capped geometries.
In principle, the functional form that this requires for both
$U_a(\phi)$ and $V_a(\phi)$ can be inferred in this way using the
known expressions for the bulk dilaton profile, $\phi(\eta_a)$,
together with the expressions for $U_a(\eta_a)$ and $V_a(\eta_a)$
obtained by solving eqs.~\pref{Isr1} and \pref{Isr2}.

Finally, the dilaton jump condition, \pref{dilatonJ2}, is then
read as an additional constraint on the parameters which govern the
capped geometry. To identify this constraint more explicitly, we
notice that we could use either the bulk dilaton profile,
$\phi(\eta)$, or the profile in the cap, $\hat\phi(\eta)$, to
convert the $\eta_a$ dependence of $U_a$ and $V_a$ into their
dependence on the dilaton. In particular, we have two ways of
evaluating the dilaton derivative of the 4-brane quantities like
$U_a$, which must agree with each other:
\be \label{ConsistentUderivative}
    \left( \frac{\exd U_a}{\exd \eta_a} \right)
    = \left( \frac{\exd U_a}{\exd
    \phi} \right)_{\phi = \phi(\eta)}
    \left( \frac{\exd \phi}{\exd \eta_a} \right)
    = \left( \frac{\exd U_a}{\exd
    \phi} \right)_{\phi = \hat\phi(\eta)}
    \left( \frac{\exd \hat\phi}{\exd
    \eta_a} \right) \,.
\ee
Here $\exd \phi/\exd \eta_a = (\partial\phi/\partial \eta)|_{\eta
\to \eta_a}$, while $\exd \hat \phi/\exd \eta_a$ also includes the
implicit dependence on $\eta_a$ that that $\hat\phi$ acquires
through its dependence on the $\eta_a$-dependent cap parameters.
Collectively denoting these cap parameters by $\{ \hat c_s \} =
\{\lambda_a, \xi_{1a},\dots \}$, we have
\be \label{ChainRule}
    \frac{\exd \hat\phi}{\exd \eta} = \left[\left( \frac{\partial
    \hat\phi}{\partial \eta} \right) + \left(
    \frac{ \partial \hat\phi}{\partial \,\hat c_s} \right)
    \frac{\partial \,
    \hat c_s}{\partial \eta_a} \right]_{\eta \to \eta_a}\,.
\ee
The desired consistency condition on the cap parameters comes from
equating $(\partial \hat\phi/\partial \eta)_{\eta \to \eta_a}$
obtained by solving eqs.~\pref{ConsistentUderivative} and
\pref{ChainRule}, with that inferred from the dilaton jump
condition, eq.~\pref{dilatonJ2}.

We see from this that the number of independent constraints on the
cap geometry is the same as it was when we made the simpler
assumption that $U$ and $V$ were constants. We have not yet tried
to solve these constraints to determine the functional form for
$U_a(\phi)$ and $V_a(\phi)$ which would be obtained.

\subsubsection*{Solving the matching conditions}
\label{solvconcon}

In order to see in more detail if obstructions to solutions to the
matching conditions might exist, we next examine some of these
conditions in more detail. Recall the counting: each cap is
described by 7 integration constants: $\la$, $q_a$, $\xi_{1a}$,
$\xi_{2a}$, $\omega_a$, $p_a$ and $\nab$, if the smoothness
condition is not used, for a total of 14 once both branes are
included. Smoothness of the caps and continuity at both branes,
with the topological condition cut these down by a total of 9
conditions, leaving 5 undetermined. There is also one combination
of jump conditions at each brane which does not involve the
potentials $U$ and $V$, reducing us to 3 parameters. If $U$ and
$V$ are $\phi$-independent, then the dilaton jump condition for
each brane removes 2 more. This leaves 1 cap parameter
undetermined. By contrast, the integers $k_a$, $k_b$ and $N$
describing the monopole flux and background configuration for the
Stueckelberg field are not solved for, but are instead regarded as
choices we get to pick by hand. We show there is a solution to the
junction conditions for a range of $k_a$, $k_b$ and $N$.

\bigskip\noindent{\it A special case:}\medskip

\noindent Before examining the general case, we first examine in
detail a special case where all of the conditions may be
explicitly solved in closed form. In order to do this, we make the
following {\em ansatz} for the integration constants, $\lambda_a$
and $\lambda_b$:
\be
    \frac{\lambda_a}{q_a}=
    \frac{\lambda_1}{q}=
    \frac{\lambda_b}{q_b} \,,\label{parchoi1}
\ee
Then, we choose $\omega_a$ and $\omega_b$ to satisfy

\be\label{parchoi2}
    \omega-\omega_a+\lambda_3 \eta_a\,=\,0\,=\,
    \omega-\omega_b+\lambda_3 \eta_b
\ee
while the parameters $q_a$ and $q_b$ are chosen such that

\be
    \frac{q_a}{g_a}\,=\,\frac{q \lambda_2}{g \lambda_1}\,=\,
    \frac{q_b}{g_b}\label{parchoi3}
\ee
The motivation for these choices comes from the way they simplify
the continuity equations.

Eq.~\pref{parchoi1} ensures that the continuity relation,
eq.~(\ref{continuity1}), simplifies to
\be
    \lambda_a \left(\eta_a-\xi_{1a}\right) = \lambda_1
    \left(\eta_a-\xi_1\right) \,, \label{consconst}
\ee
which we solve for $\xi_{1a}$, giving
\be \label{xi1aspecialcase}
    \xi_{1 a}= \eta_a - \frac{\lambda_1}{\lambda_a}
    (\eta_a - \xi_1) \,.
\ee
Similarly, eqs.~\pref{parchoi2} and \pref{parchoi3} allow the
continuity relation \pref{continuity2} to be written
\be
    \lambda_a \left(\eta_a-\xi_{2a}\right) = \lambda_2
    \left(\eta_a-\xi_2\right) \,, \label{consconst2}
\ee
with solution
\be \label{xi2aspecialcase}
    \xi_{2a}= \eta_a - \frac{\lambda_2}{\lambda_a}
    (\eta_a - \xi_2) \,.
\ee
Similar results follow for $\xi_{1b}$ and $\xi_{2b}$ using
identical arguments.

Given these conditions, the topological constraint,
(\ref{genqua}), degenerates into
\be
 \frac{N}{e} \,=\, \frac{2 \l_1}{q} \,, \label{orqua}
\ee
which is independent of the brane positions, and so can be
regarded as a condition on the background field gauge coupling,
$e$ (which can be altered by adjusting how the background gauge
field is embedded into the gauge group). Similarly, using the
choices \pref{parchoi2} and \pref{parchoi3} in
\pref{l_constraint}, derived from the jump conditions, leads to
the considerably simpler form
\be \label{l_constraint3}
    \frac{2 k_a}{N} = 1 +
    \frac{(\l_3/\l_1)}{2 \, [1-(q_a/q)]}  \,,
\ee
with a similar result for brane $b$. For $\lambda_3=0$ this last
formula requires $N$ to be even, and was obtained previously for
non-supersymmetric 6D models in ref.~\cite{OtherCap}. If
$\lambda_3 \neq 0$, on the other hand, it instead can be read as
giving ${q_a}/{q}$ in terms of $\lambda_3$. Identical
considerations similarly apply to brane $b$. Due to the condition
(\ref{parchoi3}), the condition (\ref{l_constraint3}) allows to
obtain a constraint that the parameter $g_a$ must satisfy in order
to get a solution:
\be \label{l_constraint3bis}
    \frac{2 k_a}{N} = 1 +
    \frac{(\l_3/\l_1)}{2 \, [1-{g_a \lambda_2}/{(g \lambda_1)}]}.
\ee

Next, given assumption \pref{parchoi1}, the smoothness condition,
eq.~\pref{constfcap}, reduces to
\be \label{newconstfcap}
    2 g_a e^{\lambda_a\left(\xi_{2 a}-\xi_{1 a}\right)}=
    \frac{q}{\lambda_1}=2 g_b
    e^{\lambda_b\left(\xi_{1 b}-\xi_{2 b} \right)} \,,
\ee
which, using eqs.~\pref{xi1aspecialcase} and
\pref{xi2aspecialcase}, can be reformulated as
\be \label{newconstfcapbis}
    e^{\left(\lambda_1- \lambda_2\right) \eta_a
    + \lambda_2 \xi_2
    -\lambda_1 \xi_1} =
    \frac{q}{2\lambda_1 g_a} \,.
\ee
This may be regarded as the condition that determines the brane
position $\eta_a$. Notice that this last expression, together with
its counterpart for brane $b$, gives the following constraint
relating the positions of the two branes:
\be
    \left(\lambda_1-\lambda_2 \right) \,
    \left( \eta_a-\eta_b \right)
    = \ln{ \left(\frac{q^2}{ 4
    \lambda_1^2\, g_a g_b } \right)} \,.\label{branpos}
\ee
The final parameter, $p_a$, is fixed by eq.~\pref{continuity3} to
be $p_a = \lambda_3 \eta_a$.

Finally, we solve the dilaton jump condition and the two Israel
junction conditions, which involve the 4-brane parameters $U$,
$V$, $\exd U/\exd\phi$ and $\exd V/\exd\phi$. Solving the two
Israel conditions gives the following expressions for $U_a$ and
$V_a$:
\begin{eqnarray}
    -2 \cW^4 \cA\, V_a &=& \frac{\lambda_3}{2}
    +2 \left( \lambda_a -\lambda_2 \right)
    \tanh{\lambda_2\left(\eta_a-\xi_2\right)} +
    \left( \lambda_1 -\lambda_a \right)
    \tanh{\lambda_1\left(\eta_a-\xi_1\right)}
    \nonumber \\
    -\frac{\cW^4}{\cA}\,\left(k_a-e A_{\psi}\right)^2
    \, U_a &=& \frac{\lambda_3}{2}
    +\left( \lambda_a -\lambda_1 \right)
    \tanh{\lambda_1\left(\eta_a-\xi_1\right)}\,.
\end{eqnarray}
The dilaton matching condition similarly becomes
\begin{eqnarray}
    2\lambda_3 -\left( \lambda_a -\lambda_1 \right)
    \tanh{\lambda_1\left(\eta_a-\xi_1\right)} +
    \left(\lambda_a-\lambda_2 \right) \tanh{\lambda_2
    \left(\eta_a-\xi_2 \right)} \,=\,
    \cF \left(\frac{d U_a}{d \phi} ,
    \frac{d V_a}{d \phi} \right)\label{dilmatspec}
\end{eqnarray}
where the function $\cF$ denotes the combination of the $U$ and
$V$ and their derivatives appearing on the right-hand-side of
\pref{dilatonJ} (and so $\cF = 0$, in particular, if $\exd
U_a/\exd \phi = \exd V_a/\exd \phi = 0$).

As usual, whether this last equation must be read as a new
constraint depends on the functional form which is assumed for
$U_a(\phi)$ and $V_a(\phi)$. In particular, if $U_a$ and $V_a$ are
constants (or scale invariant), then eq.~\pref{dilmatspec} imposes
non-trivial additional conditions on the parameters of the cap
geometries, and so generically can obstruct the existence of a cap
geometry unless the bulk parameters are tuned to assure its
satisfaction.

Notice that the necessity to tune parameters in the bulk and cap
actions arises in this case because the initial simplifying {\em
ans\"atze}, eqs.~\pref{parchoi1}, \pref{parchoi2} and
\pref{parchoi3}, make the matching problem into an over-determined
problem, rather than allowing the 1-parameter family of solutions
which are possible in the generic case.

\bigskip\noindent{\it The general case:}\medskip

\noindent We now return to solving the matching condition in the
general case, not subject to the {\em ans\"atze},
eqs.~\pref{parchoi1}, \pref{parchoi2} and \pref{parchoi3}. It is
convenient to define first the quantities
\bea
    && \Lambda_{ia} = \lambda_{a}(\nab-\xi_{ia}), \qquad
    \Lambda_{ib} = \lambda_{b}(\eta_b -\xi_{ib}),  \label{defoflam} \\\
    {\rm and } \qquad && \Delta_{ia} = \l_i(\nab-\xi_i), \qquad
    \Delta_{ib} = \l_i(\nbb - \xi_i) \label{defofdel}
\eea
where $i = 1,2$. In our counting, the parameters $\Lambda_{ia}$
and $\Lambda_{ib}$ replace $\xi_{ia}$ and $\xi_{ib}$, whereas
$\Delta_{ia}$ and $\Delta_{ib}$ are known functions of $\eta_a$
and $\eta_b$.

Recall that there are a total of 14 cap parameters, and these are
subject to a total of 11 conditions before the three conditions
(per brane) involving $U$ and $V$ are used, leaving 3 parameters
undetermined. (Depending on what we assume about the 4-brane
action -- such as if $U$ and $V$ are constants -- two of these can
then be fixed by the dilaton jump conditions, leaving the single
undetermined parameter, although we do not yet apply this
constraint in this section.) Although other choices are possible,
we find it easiest to solve for the cap parameters as functions of
the three undetermined quantities $(\nab,\nbb, \Lambda_{1b})$.

We start with the topological constraint, eq.~\pref{genqua}, which
we simplify by using eq.~\pref{continuity1} and its counterpart
for brane $b$ to eliminate the combinations $\l_a/q_a$ and
$\l_b/q_b$. Using the resulting expressions in eq.~\pref{genqua}
gives
\be
   \label{newtop}
   \tanh\Delta_{1b}-\tanh\Delta_{1a}
   = \frac{qN}{e\l_1}
   - \frac{ \varepsilon_a \, e^{\Lambda_{1a}} }{\cosh\Delta_{1a}}
   - \frac{    \varepsilon_b \, e^{-\Lambda_{1b}  }  }{
   \cosh\Delta_{1b}}\,,
\ee
where we define $\varepsilon_a = |q_a|/q_a = \sign \, q_a$, and
similarly for $\varepsilon_b$ and $\varepsilon$. Writing this as
$\varepsilon_a e^{\Lambda_{1a}} = F$, with $F =
F(\nab,\nbb,\Lambda_{1b})$ given by
\be
     F = \cosh\Delta_{1a}  \left(  \frac{qN}{e\l_1} -
     \frac{ \varepsilon_b e^{-\Lambda_{1b} } }{\cosh\Delta_{1b}}
     - \tanh\Delta_{1b}+ \tanh\Delta_{1a} \right) \,,
\ee
shows that solutions exist so long as we choose $\varepsilon_a =
\sign \, F$, and gives these solutions as
\be
   \Lambda_{1a} = \ln|F| \,.
\ee

Using the smoothness condition together with the continuity
condition, eq.~\pref{continuity1}, and the above solution for
$\Lambda_{1a}$, then gives
\bea
    \Lambda_{2a} &=& \ln\left| \frac{
    \l_1 g_a (1+F^2)}{ q \, \cosh\Delta_{1a} }\right|.
 \eea
As we have now solved for $\Lambda_{1a}$ and $\Lambda_{2a}$ in
terms of $\eta_a$, $\eta_b$, and $\Lambda_{1b}$, we do not bother
to eliminate these two parameters from future expressions.

We next solve for $\lambda_a$. Starting from
eq.~\pref{l_constraint} and using the continuity conditions to
simplify further, we arrive at the expression
\bea
   \lambda_a &=& \frac{1}{\tanh\Lambda_{1a} }
   \left( \l_1 \tanh\Delta_{1a}
   -\frac{\l_3}{2}  \right. \nn \\
   && \qquad\qquad\qquad \left. + \left[ 1 -
   \frac{\varepsilon \varepsilon_a g_a \l_2 \, \cosh\Delta_{1a} \,
   \cosh\Lambda_{2a}}{ g \l_1 \, \cosh\Delta_{2a}
   \, \cosh\Lambda_{1a}}
   \right] \left[ \frac{q k_a}{e} -
   \frac{\varepsilon_a \l_1 e^{\Lambda_{1a}} }{
   \cosh\Delta_{1a}} \right]  \right).
\eea
It is important to note that by choosing the integer $k_a$
appropriately, we can ensure $\lambda_a>0.$\footnote{One might
worry that this is no longer true if the first term in square
brackets is zero, but a little work shows that the condition for
this term being nonzero (for arbitrary $k_a$)
is equivalent to the condition $U_a \not=
0$, which we assume.}  Again, as we have solved for $\lambda_a$ in
terms of the three required parameters, we will not need to
eliminate it from future equations. Finally, the 3 continuity
equations at brane $a$ directly give
\bea
   p_a &=& \l_3 \nab, \\
   q_a &=& \left( \frac{ \varepsilon q \la}{\l_1} \right)
   \left( \frac{2F}{1+F^2} \right)
   \cosh\Delta_{1a}, \\
   \omega_a &=& \omega + \l_3 \nab + \frac{1}{2} \ln\left|
   \frac{ g_a \l_2 \, \cosh\Lambda_{2a} }
   {g \l_a \, \cosh\Delta_{2a} } \right|.
\eea
The analysis at brane $b$ is similar, for which we find
\bea
   \label{braneb1}
   \Lambda_{2b} &=& \Lambda_{1b}
   + \ln \left|  \frac{ q \, \cosh\Delta_{1b} }
   {2\l_1 g_b \, \cosh\Lambda_{1b} }   \right|, \\
   \lambda_b &=& \frac{1}{\tanh\Lambda_{1b} }
   \left( \l_1 \tanh\Delta_{1b}
   -\frac{\l_3}{2}   \right. \nn \\
   && \qquad \qquad \qquad  \left. + \left[ 1 -
   \frac{ \varepsilon \varepsilon_b  g_b \l_2 \, \cosh\Delta_{1b} \,
   \cosh\Lambda_{2b}}{ g \l_1 \, \cosh\Delta_{2b}
   \, \cosh\Lambda_{1b}}
   \right] \left[ \frac{q k_b}{e} +
   \frac{\varepsilon_b \l_1 e^{-\Lambda_{1b}} }{
   \cosh\Delta_{1b}} \right]  \right),
\eea
and
\bea
    p_b &=& \l_3 \nbb, \\
    |q_b| &=& \left( \frac{|q| \l_b}{\l_1} \right)
   \frac{\cosh\Delta_{1b} }{ \cosh\Lambda_{1b} }, \\
   \omega_b &=& \omega + \l_3 \nbb + \frac{1}{2} \ln\left|
   \frac{ g_b \l_2 \, \cosh\Lambda_{2b} }
   {g \l_b \, \cosh\Delta_{2b} } \right|.     \label{braneb2}
\eea
By using the previous expressions for $\Lambda_{2b}$ and $\l_b$,
we see that we have solved for $|q_b|$ and $\omega_b$ in terms of
the required 3 parameters. The sign of $q_b$ can be determined by
the gauge field jump condition at brane $b$.

This exhausts all of the matching conditions which do not involve
the 4-brane coupling functions. The value of these functions, $U$
and $V$, at each brane is then easily obtained by solving the
Israel junction conditions, eqs.~\pref{Isr1} and \pref{Isr2},
leaving only the dilaton jump condition to be solved. If $U$ and
$V$ contain enough parameters to allow them and their derivatives
to be varied independently for each brane, then this last
condition can be solved without adding further constraints on the
parameters of the cap geometry.

Alternatively, when $\exd U/\exd \phi$ and $\exd V/\exd \phi$ are
not independent of $U$ and $V$ --- such as when $U$ and $V$ are
both $\phi$-independent, or are scale invariant --- then the
dilaton matching condition, eq.~\pref{dilatonJ}, imposes an
additional constraint. After some manipulation this can be written
in the form
\begin{eqnarray} \label{dilatongeneral}
    && g\, \cosh{\Delta_{2a}} \,\left[
    2 \lambda_3 +\lambda_1 \tanh{\Delta_{1a}}-
    \lambda_2 \tanh{\Delta_{2a}} \right] \nonumber\\
    && \hskip 1.4cm =
    \frac{g_a^2 \,\lambda_1 \lambda_2}{2 q \, \cosh{\Delta_{1a}}}
    \left\{ \left( \frac{q^2}{\lambda_1^2 g_a^2} \right)
    \, \cosh^4{\Delta_{1a}}\,\left[\frac{\left(
    \tilde{N} - \varepsilon_b e^{-\Lambda_{1b} } \sech\Delta_{1b}
    \right)^2}{1+\left(
    \tilde{N} - \varepsilon_b e^{-\Lambda_{1b} } \sech\Delta_{1b}
    \right)^2}\right]-1 \right\} \nonumber\\&&
\end{eqnarray}
where we define the quantity
\be
    \tilde{N}\,=\,\frac{qN}{e\l_1} - \tanh\Delta_{1b}+
    \tanh\Delta_{2a} \,.
\ee
A particularly useful special of this condition takes $\eta_a$ to
be very large and negative (and $\eta_b$ to be large and
positive). This is a limit of particular interest because it
corresponds to the cap volume being much smaller than that of the
bulk (more about this in subsequent sections). In this limit we
have $\tilde{N} \approx -2 + qN/(e\lambda_1)$ and the previous
equation reduces to
\be \label{dilatonlargeeta}
    g\, \cosh{\Delta_{2a}}
    \,\left[
    2 \lambda_3 -\lambda_1 +  \lambda_2
    \right] = \left(
    \frac{ \, \lambda_2 \,q}{2\,
    \lambda_1 } \right) \,
    \cosh^3{\Delta_{1a}}\,\left[\frac{
    \left(\tilde{N} - \varepsilon_b e^{-\Lambda_{1b} }
    \sech\Delta_{1b} \right)^2}{1+\left( \tilde{N} -
    \varepsilon_b e^{-\Lambda_{1b} } \sech\Delta_{1b}
    \right)^2}\right] \,.
\ee

Recall that eq.~\pref{dilatongeneral} --- or
eq.~\pref{dilatonlargeeta} --- and its counterpart for cap `$b$'
impose two conditions on the three remaining free cap parameters,
$\eta_a$, $\eta_b$ and $\Lambda_{1b}$. In particular, in the limit
of large negative $\eta_a$ and large positive $\eta_b$, this
equation is easily solved for $\eta_a$ because $\tilde{N}$ is
independent of $\eta_a$ and $\eta_b$. In general, the freedom to
choose $N$ can be used to help ensure that solutions exist.

\subsection{Bulk geometries sourced by given branes}

In the previous section the bulk geometry is considered to be
given, and we ask whether regularizing caps can be constructed.
This section adopts a different point of view, wherein the
characteristics of the caps --- {\it i.e.} all the integration
constants that define the cap geometry and the quantities $U$ and
$V$ --- are given, and we seek the properties of the bulk which
results. In particular, our interest is to see whether and how the
two caps must be related to one another, and to check whether the
bulk configuration is always of the form of a GGP solution, with
flat four dimensional slices.

Our goal in doing so is two-fold.  First, in this subsection, we
wish to see whether this reduced problem is over-determined, and
if so what is required in detail of the branes in order to ensure
a solution. Secondly, in \S\ref{stabilznsection} we set out to
understand how the volume of the bulk geometry is related to the
brane properties, and, by doing so, to exhibit a stabilization
mechanism for the bulk volume. Of particular interest is then to
understand what 4-brane/cap properties are required to ensure the
volumes of the capped regions are much smaller than that of the
intervening bulk (as is required if the 4-branes and caps describe
the microscopic structure of more macroscopic 3-branes).

\subsubsection*{Parameter counting and junction conditions}

We now show that counting equations and parameters suggests we are
not completely free to specify the 4-brane action for brane $a$
arbitrarily if we ask that it interpolate between 4D flat cap and
bulk geometries. This can be done only if the 4-brane action is
subject to one constraint equation (as was argued in
ref.~\cite{NavSant}), but once this is satisfied there is
sufficient information to determine the parameters describing both
the bulk geometry {\it and} the properties of brane
$b$.\footnote{To be precise, we find a two parameter family of
solutions for the bulk and cap $b$, corresponding to where we
choose to embed the two branes in the bulk. Once this choice is
made, then the bulk and cap $b$ are unique.}

To this end, imagine we specify the cap geometry and 4-brane
action at a given position $\eta = \eta_a$. Next recall that there
are 7 integration constants characterizing the the bulk geometry
--- $\lambda_1$, $\lambda_2$, $\xi_1$, $\xi_2$, $p$, $q$ and
$\omega$. (Notice that, although previously we have removed two
of these quantities -- $\xi_1$ and $p$ -- by suitably
adjusting coordinates, this is typically no longer possible
without altering the specified parameters for cap $a$.) These 7
parameters are subject to a total of 7 conditions at $\eta_a$,
consisting of 3 continuity conditions (metric and dilaton) and 4
jump conditions (Israel, Maxwell and dilaton), suggesting that
the bulk parameters are completely specified in terms of those of the
cap and 4-brane at $\eta_a$.

As we show in the next section, however, one of these seven
equations which is {\it supposed} to determine one of the bulk
parameters turns into a constraint equation amongst cap and
brane parameters.
Thus, what we find is that for any given cap and brane which
satisfies the constraint, there is a one-parameter family of flat
bulks to which we can match. Physically, it is easiest to
interpret this one parameter in the coordinate system where $\xi_1
= \xi_{1a} = 0$. Recall that in this coordinate system the brane
location in the bulk and cap is $\eta_a$ and $\hat{\eta}_a$,
respectively, where these two numbers are generically not the
same. Here, we again imagine fixing the cap and brane properties
at $\hat{\eta}_a$, and then solving for six of the seven bulk
parameters: $\l_1$, $\l_2$, $\xi_2$, $\omega$, $p$, $q$, and
$\eta_a$. Thus, this one-parameter family of bulk solutions
corresponds to where we choose to place the brane in the bulk
coordinate system. If $\eta_a$ is fixed, then we find a unique
solution for the bulk.

Continuing to use the coordinate system where $\xi_1 = \xi_{1a} =
\xi_{1b} = 0$, we see that once the bulk geometry is thus
inferred, there remain 10 parameters associated with cap $b$,
consisting of 6 integration constants --- $\lambda_b$, $q_b$,
$\xi_{2b}$, $\omega_b$, $p_b$, and $\hat{\eta}_b$ --- plus the
brane position $\eta_b$ in the bulk coordinate system, the two
4-brane parameters, $U_b$ and $V_b$, and {\it one} linear
combination of their derivatives. These 10 parameters are then
subject to 9 conditions, consisting of the 7 continuity and jump
conditions at the brane location, the smoothness condition at
$\eta \to \infty$ for cap $b$ and the topological constraint on
the Maxwell field. Provided there are no obstructions to solving
these equations, this shows that once we choose the properties of
one brane (subject only to the Hamiltonian constraint), together
with the location of the two branes in the bulk coordinate system,
$\eta_a$ and $\eta_b$, then properties of the other brane and the
intervening 4D-flat bulk are precisely dictated. If the properties
of brane $b$ are not adjusted in this way in terms of those of
brane $a$ then the intervening bulk solution cannot be 4D flat,
and instead must either be 4D maximally symmetric but not flat
\cite{6DdSSUSY} or time-dependent and not Lorentz invariant
\cite{Scaling}.

This counting bears out, and make more precise, expectations based
on earlier studies of the general properties of bulk solutions to
6D supergravity. In particular, for 4D maximally-symmetric
solutions \cite{6DdSSUSY} (including those which are not 4D flat)
the bulk geometry depends nontrivially only on $\eta$, and so we
may imagine integrating the bulk field equations in the $\eta$
direction, starting at brane $a$ and ending at brane $b$. Since
the $\eta$-$\eta$ Einstein equation does not involve second
derivatives of the metric, it represents a `Hamiltonian'
constraint on those `initial' conditions at brane $a$ which can be
consistently used for such an integration. In this language, the
above-mentioned constraint on the allowed 4-brane parameters
corresponds to requirements imposed on the 4-brane by matching to
the Hamiltonian constraint in the bulk, restricted to 4D flat
geometries \cite{NavSant}. Furthermore, since the bulk geometry is
completely specified by integrating forward in $\eta$ using the
`initial' conditions at brane $a$, its asymptotic form at brane
$b$ is seen to be completely determined, in agreement with what we
find here for explicit 4-brane/cap regularizations of this
asymptotic form.

\subsubsection*{Explicit solutions}

To better see if parameter and equation counting provides the
whole story, we next solve the matching to see whether
obstructions to their solutions can exist.

\medskip\noindent{\it The Bulk}\medskip

\noindent The continuity equations, eqs.~\pref{continuity1} --
\pref{continuity3}, read in this case:
\bea
   |q| &=& |q_a| \left( \frac{\l_1 \, \cosh\Lambda_{1a}}{
   \la \, \cosh\Delta_{1a}} \right) \,, \\
   e^{-2\omega} &=& e^{-2(\omega_a-\l_3 \nab)}
   \left( \frac{g_a \l_2 \, \cosh\Lambda_{2a}}{g
   \la \, \cosh\Delta_{2a}} \right) \,, \label{omega} \\
   p &=& p_a -\l_3 \nab\,, \label{p}
\eea
and can be thought as fixing the bulk parameters $q$, $\omega$,
and $p$ (recall the definitions of the parameters $\Lambda$ and
$\Delta$ in formulae (\ref{defoflam}) and (\ref{defofdel})). Note
that the sign of $q$ is not yet fixed. These solutions are given
in terms of the four bulk quantities $\l_1$, $\l_2$,
$\Delta_{1a}$, and $\Delta_{2a}$, for which we now solve.

Before proceeding it is convenient to first define four
combinations of brane and cap parameters:
\bea
    \cC_1 &=& \left( \frac{g_a \cosh\Lambda_{2a} }{2 g \la}
    \right)  \left[ \hcA \hcW^4(V_a-2V_a') -
    \frac{\hcW^4}{2\hcA}(U_a + 2U_a')(k_a-e\hat{A}_{\psi})^2
    \right],\label{defC1} \\
    \cC_2 &=&  \left( \frac{g_a \cosh\Lambda_{2a} }{4 g \la} \right)
     \left[ \hcA \hcW^4 (5V_a-2V_a')  \right. \nn \\
    && \left. \qquad \qquad \qquad +
    \frac{\hcW^4}{2\hcA}(3U_a-2U_a')(k_a-e\hat{A}_{\psi})^2
    +4\la \tanh\Lambda_{2a} \right] ,\\
    \cC_3 &=& \left( \frac{ \varepsilon_a g_a
    \cosh\Lambda_{2a}}{ g \cosh\Lambda_{1a} } \right)
    \left[ -\frac{e  U_a }{q_a}  \left(\frac{\hcW^4}{\hcA}\right)
    (k_a - e\hat{A}_{\psi}) + 1 \right], \\
    \cC_4 &=&   \left( \frac{g_a \cosh\Lambda_{2a} }{4 g \la} \right)
    \left[ \hcA \hcW^4 (V_a-2V_a') \right. \nn \\
     && \left. \qquad \qquad \qquad  +
     \frac{\hcW^4}{2\hcA}(7U_a-2U_a')(k_a-e\hat{A}_{\psi})^2
     + 4\la \tanh\Lambda_{1a}
    \right] \,,\label{defC4}
 \eea
where primes here denote differentiation with respect to $\phi$.
These four parameters will take the place of $U_a(\nab)$,
$V_a(\nab)$, their derivatives (which appear in only one linear
combination), and $k_a$. The action for brane $a$ can therefore be
equally well characterized by these four quantities, as by our
original parameterization in terms of $U_a(\nab)$, $V_a(\nab)$,
and derivatives. With these definitions in hand, the remaining
four matching conditions reduce to the following equations:
\bea
   \cC_1 &=& \cosh\Delta_{2a} \left[ 1 -
   \left(\frac{\l_1}{\l_2}\right)
   \right]^{\frac{1}{2}} \label{cap2bulk_1}\\
   \cC_2 &=& \sinh\Delta_{2a}  \label{cap2bulk_2} \\
   \cC_3 &=& \varepsilon \left(\frac{\l_1}{\l_2} \right)
   \frac{\cosh\Delta_{2a}}{\cosh\Delta_{1a}}  \label{cap2bulk_3} \\
   \cC_4 &=& \left(\frac{\l_1}{\l_2}\right)
   \tanh\Delta_{1a} \cosh\Delta_{2a}  \label{cap2bulk_4}.
\eea
Recalling that both $\l_1$ and $\l_2$ are positive, we see
immediately that $\varepsilon \equiv \sign \, q = \sign \, \cC_3$.

We note that this system of equations is over-determined, since
there are four equations but only three unknowns: $\Delta_{1a}$,
$\Delta_{2a}$, and $\l_1/\l_2$. In fact, by squaring the above
equations it is straightforward to check this constraint is given
by
\be
   \label{hamconst}
   \cC_1^2 - \cC_2^2 + \cC_3^2 + \cC_4^2 = 1.
\ee
When the above equation is satisfied, then it can be shown that
the bulk fields satisfy the Hamiltonian constraint which ensures
4D flatness.\footnote{This Hamiltonian constraint is given by eq.
(34) in reference \cite{6DdSSUSY}.} Henceforth, we assume that the
brane properties are chosen such that the Hamiltonian constraint
is satisfied. In this case, the solution to eqs.~\pref{cap2bulk_1}
- \pref{cap2bulk_4} is
\bea
   \Delta_{1a} &=& \sign( \cC_4 ) \, \arcosh\left[ \left( 1 +
   \frac{\cC_4^2}{\cC_3^2} \right)^{\frac{1}{2}}  \right], \\
   \Delta_{2a} &=& \arsinh( \cC_2 ), \\
   \frac{\l_1}{\l_2} &=& \left( 1 - \frac{\cC_1^2}{\cC_1^2 + \cC_3^2 + \cC_4^2} \right)^{\frac{1}{2}}.
\eea
where the range of arcosh is taken be $\{ x \in {\mathbb R}: x \ge 0 \}$. It is easy to see that solutions to these equations exist for any values of the $\cC_i$, subject only to the constraint that they obey eq.~\pref{hamconst}.

As expected from the arguments in the previous section, we indeed
find a one-parameter family of possible bulks. Once this parameter
is fixed --- corresponding to choosing where in the bulk we wish
to embed the brane -- then the bulk solution becomes unique.
Henceforth, we assume that this choice has been made (as can be
accomplished by making a specific choice for $p$ in eq.~\pref{p}).

\medskip\noindent{\it Cap $b$}\medskip

\noindent Having uniquely determined the bulk solution, it remains
to determine the properties of the 4-brane and cap at brane $b$.
In order to find a unique solution, we first specify the location
where we wish to cap the bulk, $\eta_b$. Since this analysis is
identical to that of \S\ref{cappingbulk}, we do not repeat it in
detail here, however the three continuity conditions, the
smoothness condition, and the combination of the jump conditions
which is independent of $U_b$ and $V_b$ provide 5 constraints on
the 5 cap integration constants $p_b$, $\Lambda_{2b}$,
$\lambda_b$, $q_b$ and $\omega_b$ (see eqs.~\pref{braneb1} -
\pref{braneb2}). Then, the two Israel junction conditions fix
$U_b$ and $V_b$, and the dilaton jump condition provides the
constraint which fixes the one relevant combination of derivatives
$U_b'$ and $V_b'$. The only cap parameter which is not fixed by
these conditions is $\Lambda_{1b}$, and this can be determined
from the topological equation \pref{newtop}.
As expected \cite{6DdSSUSY}, both the properties of the bulk and
those of the 4-brane and cap at $\eta = \eta_b$ are dictated by
those of the brane and cap at $\eta = \eta_a$.

\subsection{Volume stabilization and large hierarchy}
\label{stabilznsection}

The previous analysis fixing the seven bulk integration constants
in terms of given cap parameters fixes in particular the
integration constant, $\omega$, that parameterizes the bulk
volume. This provides a natural 6D mechanism for stabilizing this
bulk volume. In this section we explore this stabilization in more
detail, focussing on the conditions which are required to obtain a
large hierarchy between the volumes of the bulk and the caps. In
the next section we identify the low-energy 4D effective potential
which is generated in this way for $\omega$.

\subsubsection*{Conditions for a hierarchy}

We now ask for the conditions the brane and cap actions should
satisfy to ensure that the cap volumes are much smaller than those
of the bulk. Our point of view here is that the bulk geometry is
given, and so would like to phrase the conditions for a hierarchy
in terms of only those parameters over which we have control: the
bulk parameters and the three cap parameters, $\nab$, $\nbb$, and
$\Lambda_{1b}$.

In order to have branes whose circumference is small, we seek to
ensure $\cA(\nab)$ and $\cA(\nbb)$ are much less than
one. We see from eq.~\pref{eom2} that it is natural to
examine for this purpose the limit $\nab \rightarrow -\infty$ and
$\nbb \rightarrow \infty$, although in general this need not be
sufficient in itself to have small cap volumes. However, we
now argue that sufficient conditions for obtaining small cap
volumes are given by
\bea
    \label{hierarchy}
   \Lambda_{1a} =  \l_a(\nab - \xi_{1a}) &\ll& -1 \nn  \\
   \Lambda_{2a} = \l_a(\nab-\xi_{2a}) &\ll& -1
\eea
with similar conditions for brane $b$. Large, negative $\nab$ is
not sufficient for small cap volumes because it does not in itself
ensure that these conditions are satisfied. Under these conditions
we may use the asymptotic form for the hyperbolic functions and so
obtain the following expression for volume of cap $a$
\bea
   \label{Omega}
   \Omega_a &=& 2\pi \int_{-\infty}^{\eta_a} \exd \eta \;
    \hcA^2 \hcW^4 \nonumber\\
   &\simeq& \frac{\pi}{\l_a} \,e^{2(\omega-\omega_a+p_a)} (\cA^2 \cW^4)|_{\nab}
\eea
In arriving at the second line we have used the continuity equations, \pref{J1}, to relate cap functions to bulk functions. 
The cap volume must be compared with  the bulk
volume, given by the expression
\bea
 \label{bulkvolume}
 \Omega_{bulk} &=& 2\pi \int_{\eta_a}^{\eta_b} \exd \eta \;
    A^2  W^4 \nonumber\\
   &=& \left( \frac{ \left(2 \pi\right)^2 \l_1 \l_2^{3} \, e^{2\omega}}{
   \left(2 g\right)^3 q } \right)^{\frac12}
       \int_{\eta_a}^{\eta_b} \exd \eta \;
\frac{e^{\l_3 \eta}}{\cosh^{\frac32}{\left[ \l_2 \left(\eta-\xi_2\right)\right]}
\cosh^{\frac12}{\left[ \l_1 \left(\eta-\xi_1\right)\right]}
}\,.
\eea
It is simple to check that the integral in the previous expression
is always finite. Then, it is enough to choose the parameters in the
bulk of order one, to obtain  $\Omega_{bulk} \simeq {\cal O}(1)$.
In order to obtain a hierarchy between bulk and cap volumes, it
is necessary to demand that  $\Omega_{a} \ll {\cal O}(1)$.

\smallskip

At this point we  divide our discussion into two parts: first we consider the `special
case' cap solution discussed in section \S\ref{cappingbulk}, whose
simplicity allows simple explicit solutions. We then discuss the
same question in the more general case.

\medskip\noindent {\it The special case}\medskip

\noindent Using the solutions found in the `special case' section together with
the continuity equation \pref{continuity2}, we may evaluate
the cap volume, eq.~\pref{Omega}, in terms of bulk parameters:
\bea
    \Omega_a &\simeq& 2 \pi \left( \frac{4\lambda_2
    \lambda_1^4 \,g_a}{g\,q^4} \right)^{1/2} \,
    \exp \Bigl[\omega -2 \lambda_1 \xi_1 + \left( \lambda_3
    +2\lambda_1\right)\eta_a \Bigr] \,.
    \label{volumspec}
\eea
Now, since the coefficient of $\eta_a$ in these expressions is positive, it is
clear that taking $\eta_a$ large and negative corresponds here to
making $\Omega_a$ small. Also, from eqs.~\pref{consconst} and \pref{consconst2}
we see that the hierarchy assumptions, eq.~\pref{hierarchy}, are easily satisfied
in the limit we consider. Thus, we were indeed justified in using the asymptotic
form for the hyperbolic functions.

But for the assumptions of the `special case' model, matching also
gives the value of $\eta_a$ as
\be
    (\lambda_2-\lambda_1) \eta_a =
    \lambda_1 \xi_1-\lambda_2 \xi_2
    + \ln \left( \frac{2 \lambda_1 g_a}{q} \right) \,,
\ee
which shows that $\eta_a$ can be made large and negative if we
take $g_a$ to be small. Then equation (\ref{l_constraint3bis})
shows that this condition can be achieved by choosing the bulk
parameters such that
\be
    \frac{2 k_a}{N} \simeq 1 +
    \frac{\lambda_3}{2 \lambda_1} \,.
\ee
If this condition is satisfied, then the volume of cap $a$ is
small. Analogous considerations for cap $b$ give similar results.

\medskip\noindent {\it The general case}\medskip

\noindent We now evaluate the cap volume using the general
solutions found earlier. If we also use the hierarchy assumptions,
eq.~\pref{hierarchy}, and the continuity equation
\pref{continuity2}, we calculate the cap volume to be
\bea
   \Omega_a  &\simeq& 2\pi \left( \frac{ g \, \l_1 \,
   \cosh\Delta_{2a} }{ |q| \l_2 \, \cosh\Delta_{1a} }
   \right) \, (\cA^2 \cW^4)|_{\nab} \nn \\
   &=& \pi \cA^2|_{\nab} \,.
\eea
For the generic situation of $\O(1)$ bulk parameters, we see from
eq.~\pref{eom2} that $\cA^2|_{\nab} \ll 1$ in the limit of large
$|\nab|$. Thus, we obtain the desired result: $\Omega_a \ll
\Omega_{bulk} \sim \O(1)$. Alternatively, 
if we instead wish to have cap volumes
which are $\O(1)$ and bulk volumes which are much larger, we
simply need to choose $\omega \gg 1$ while keeping all other bulk
parameters fixed.

It remains now to show what conditions must be imposed on the bulk
parameters and cap parameters in order to ensure that conditions
\pref{hierarchy} are satisfied. To simplify this discussion, we
only consider the case $\lambda_3 =0$. We accomplish this by
adjusting the background gauge coupling, $e$, so that it is
approximately equal to its value, $e_0 = qN/(2\l_1)$, in the
absence of caps. More precisely, if we define
\be
   \epsilon = \frac{1}{2} e^{-\l_1(\nab-\xi_{1})}
   \left[ \frac{qN}{\l_1 e}-2 \right] \,,
\ee
then we should take
\be
    \epsilon \ll 1  \qquad {\rm and } \qquad \Lambda_{1b} \gg 1
\ee
and, for definiteness, take $\nab \approx -\nbb$. In this case,
the general cap solutions found earlier satisfy the desired
hierarchy conditions \pref{hierarchy}. The analogous hierarchy
conditions at brane $b$ are much simpler to satisfy due to the
fact that we get to choose freely $\Lambda_{1b}$. For example,
choosing $\nbb$ large and $\Lambda_{1b} \sim \Delta_{1b} \gg 1$
guarantees that $\Lambda_{2b}  \gg 1$ and so the two
hierarchy constraints are satisfied.

To summarize, we see here how to obtain regularizing caps which
are much smaller than the bulk volume, by appropriately tuning the
gauge coupling $e$ and by choosing large coordinate values for the
brane positions. We have also shown that requiring such a
hierarchy at only a single brane is not difficult to achieve in
the sense that it involves no tuning of any bulk parameters.

\subsection{Low-energy 4D effective potential}

We next dimensionally reduce the capped bulk to 4 dimensions in
order to identify more explicitly how 4-brane action influences
the stabilization of the would-be flat direction parameterized by
$\omega$. In this section we restrict ourselves to evaluating the
effective 4D potential for $\omega$ within the classical
approximation.

To this end we identify the effective 4D action $S_{\rm eff} =
\int \exd^4x \; {\cal L}_{\rm eff}$ by computing the 6D action at
a one-parameter family of classical solutions labelled by the
constant $\omega$:
\be \label{4DEFTdef}
    S_{\rm eff} = {\cal S}_{a} + {\cal S}_{b}
    + S_{{\rm cap}\,a} + S_{{\rm bulk}} + S_{{\rm cap}\,b} \,,
\ee
where ${\mathcal S}_a = \int \exd^4x \; {\cal L}_a$ and ${\mathcal
S}_b = \int \exd^4x \; {\cal L}_b$ represent the 4-brane action
for caps $a$ and $b$, given by eq.~\pref{4brane}, while $S_M =
\int_M \exd^6x \, {\cal L} + S_{GH}(\partial M)$ represents the 6D
bulk action, including the Gibbons-Hawking boundary contribution,
defined by eqs.~\pref{6DSugraAction} and \pref{GHTerm}. The three
last terms correspond to dividing the integration over the 2 extra
dimensions into the three intervals defining the bulk, cap $a$ or
cap $b$.

Following \cite{SLED2}, we see that using the 6D field equations,
\pref{fieldequations}, to simplify the 6D bulk action in a region
$M$ with boundaries leads to the simple expression (with $\kappa^2
= 1$)
\be
    S_{\rm cl} = \frac{1}{2} \int_M \exd^6x\,
    \sqrt{-g} \, \Box \phi_{\rm cl} -
    \int_{\partial M} \exd^5x \, \sqrt{-\gamma}
    \, K_{\rm cl} \,,
\ee
which, together with Gauss' Law, allows the last three terms in
eq.~\pref{4DEFTdef} to be written
\be
    S_{{\rm cap}\,a} + S_{{\rm bulk}} + S_{{\rm cap}\,b}
    = - \frac{1}{2} \int \exd^5x \,\Bigl( [\sqrt{-g}\,
    \partial^\eta \phi + 2 \sqrt{-\gamma}\,K]_{\eta_a}
    + [\sqrt{-g}\, \partial^\eta \phi + 2 \sqrt{-\gamma} \,
    K]_{\eta_b} \Bigr) \,,
\ee
where as before $[f(\eta)]_{\eta_a} = f(\eta_a + \epsilon) -
f(\eta_a - \epsilon)$ (and similarly for $\eta_b$).

Writing $S_{\rm eff} = \int \exd^4x \, {\cal L}_{\rm eff}$, and
evaluating the right-hand-side of this last expression using the
Israel and dilaton jump conditions, \pref{Isr1}, \pref{Isr2} and
\pref{dilatonJ} finally gives
\bea
    {\cal L}_{\rm eff} &=& 2\pi \sum_{i=\,a,b}
    \cA \cW^4 e^{2(\omega-p)} \left[
    \left( - V_i + \frac{5V_i}{4}
    - \frac12 \, \frac{\exd V_i}{\exd \phi} \right)
    + \frac{1}{2\cA^2}
    \left( - U_i + \frac{3U_i}{4}
    - \frac12 \, \frac{\exd U_i}{\exd \phi} \right)
    (k_i - eA_\psi)^2 \right] \nn\\
    &=& \pi \sum_{i=\,a,b}
    \cA \cW^4 e^{2(\omega-p)} \left[
    \left( \frac{V_i}{2} - \frac{\exd V_i}{\exd \phi} \right)
    - \frac{1}{2\cA^2}
    \left( \frac{U_i}{2} + \frac{\exd U_i}{\exd \phi} \right)
    (k_i - eA_\psi)^2 \right]  \,.
\eea
Finally, to make the $\omega$-dependence explicit we write $\cA =
\cA_0 e^{\omega/2}$, $\phi = \phi_0 - \omega$, and choose for
concreteness $V(\phi) = v \, e^{s \,\phi}$ and $U(\phi) = u \,
e^{t\,\phi}$. Identifying $V_{\rm eff} = - {\cal L}_{\rm eff}$, we
find
\be
    V_{\rm eff}(\omega) = \sum_{i=\,a,b} \left[ C_{Vi}
    \, e^{(5/2 - s_i)\,\omega} +
    C_{Ui} \, e^{(3/2-t_i) \,\omega} \right]  \,,
\ee
where
\bea \label{CResults}
    C_{Vi} &=&
    \pi \left[ \cA_0 \cW^4
    \left(\frac12 - s_i \right) v_i \, e^{s_i\phi_0-2p}
    \right]_{\eta=\eta_i}
    \nn\\
    C_{Ui} &=& - \frac{\pi}{2}\left[ \frac{\cW^4}{\cA_0}
    \left(\frac12 + t_i \right) u_i \, e^{t_i\phi_0-2p}
    (k_i - eA_\psi)^2 \right]_{\eta=\eta_i} \,.
\eea

It is clear that this potential generically only has runaway
solutions when both $C_{Ui}$ and $C_{Vi}$ and both of the
coefficients of $\omega$ in the exponents have the same sign, but
has nontrivial minima when some of these signs differ. Given the
explicit relative sign appearing in eqs.~\pref{CResults}, and
positive $u_i$ and $v_i$, we expect that stabilization of $\omega$
to be fairly generic.

\subsubsection*{The Scale Invariant Case}

Of particular interest is the case of scale-invariant branes, for
which we have $s_i = 1/2$ and $t_i = -1/2$. In this case, not only
do we recover the generic scale-invariant form for the potential
\be
    V_{\rm eff}(\omega) = C \, e^{2\omega}
    \quad \hbox{with} \quad
    C = \sum_{i= \,a,b}\Bigl( C_{Ui} + C_{Vi} \Bigr) \,,
\ee
but we also learn that $C = C_{Ui} = C_{Vi} = 0$. This agrees, and
makes more precise, the arguments of ref.~\cite{SLED2}, wherein
the same conclusion was drawn when scale-invariant branes were
characterized as delta-function sources.

\section{Conclusions}

In this paper we present a regularization procedure for resolving
the singularities in the most general axially symmetric, 4D-flat
solutions to 6D gauged, chiral supergravity. This procedure
resolves the singularities of these geometries using an explicit,
but broad, class of cylindrical 4-branes that couple with the bulk
Maxwell, dilaton and gravitational fields. The space interior to
these 4-branes is capped off using the most general smooth,
4D-flat, and axially symmetric solutions to the same 6D
supergravity equations that were used in the bulk between the two
branes. Our analysis provides the necessary tools required to
precisely explore the connections between properties of the bulk
field configurations and the structure of the branes which source
them.

We keep our analysis very general, with the goal of being able to
map out these connections with as few restrictions as possible. We
show, in particular, that the class of caps and 4-brane actions we
consider contain sufficient numbers of parameters to cap an
arbitrary axially-symmetric and 4D-flat bulk geometry. We also
show that once the properties of one of the 4-brane caps is
specified, there are sufficient parameters in the bulk geometry
and in the other cap  to complete the geometry. This both
identifies the properties of the bulk sourced by a given brane,
and precisely identifies how the properties of the brane at the
other end of the bulk are dictated by those of the source brane
with which one starts.

Knowing the properties of the caps shows that the presence of
regularizing branes has important consequences on the properties
of the bulk solutions. In particular, we show how the classical
degeneracy amongst bulk geometries having different volumes can be
lifted by the coupling of the 4-branes with the 6D dilaton. This
provides a stabilization mechanism for the bulk, which relates the
size of the extra dimensions with brane properties. By performing
a dimensional reduction we also identify the effective 4D
potential which captures this stabilization mechanism in the
low-energy limit. We are able to do because our regulated 6D
configurations are smooth everywhere, with the bulk fields not
diverging at the brane positions (as they do for the effective
co-dimension two 3-branes obtained in the thin-brane limit when
the circumference of the 4-brane is taken to zero).

There are several directions in which our geometrical construction
of the regularizing caps might be extended. First, the form of
4-brane action considered could be further generalized, such as by
depending on additional brane-localized fields. The back-reaction
of such fields on the bulk configuration could then be
consistently taken into account by studying their effects on the
continuity and junction conditions. An important special case
along these lines consists of studying the effects of integrating
out massive brane fields, to see how this affects the condition of
4D flatness. Work along these lines is currently in progress
\cite{Inprogress}.

Alternatively, our analysis could also be extended by generalizing
the class of bulk configurations for which caps can be
constructed. Of particular interest is such an extension to bulk
geometries which are not 4D flat \cite{6DdSSUSY}, for which one
might imagine using regulating cap geometries which are less
symmetric than the ones we consider here. Alternatively,
extensions to configurations in more than six dimensions are also
of interest, since bulk fields generically diverge at brane
positions in this case as well.

Such  constructions would be particularly useful for identifying
more precisely how the cosmological constant problem gets
rephrased in its extra-dimensional context. For these purposes it
is important to be able to find regularizing caps that are general
enough to characterize a large class of bulk geometries, in order
to explore all of the naturalness issues which might be associated
with a given regularization procedure.

\section*{Acknowledgements}

We are pleased to thank Claudia de Rham and Andrew Tolley for many
helpful discussions and insights. CB and DH are supported in part
by funds from Natural Sciences and Engineering Research Council of
Canada, and CB also acknowledges the Killam Foundation, McMaster
University and Perimeter Institute for research support. GT is
partially supported by the EC $6^{th}$ Framework Programme
MRTN-CT-2004-503369, and  by the EU 6th Framework Marie Curie
Research and Training network ``UniverseNet''
(MRTN-CT-2006-035863).


\begin{thebibliography}{99}

\bibitem{SS}
A.~Salam and E.~Sezgin, ``Chiral Compactification On Minkowski
$\times S^2$ Of N=2 Einstein-Maxwell Supergravity In
Six-Dimensions,'' Phys.\ Lett.\ B {\bf 147} (1984) 47;
%
S. Randjbar-Daemi, A. Salam, E. Sezgin and J. Strathdee, {\it
Phys. Lett.} {\bf B151} (1985) 351.

\bibitem{NS}
H. Nishino and E. Sezgin, {\it Phys. Lett.} {\bf 144B} (1984) 187;
``The Complete N=2, D = 6 Supergravity With Matter And Yang-Mills
Couplings,'' Nucl.\ Phys.\ {\bf B278} (1986) 353.

\bibitem{6DSugra}
Other 6D supergravities are discussed in
%
  N. Marcus and J.H. Schwarz, Phys.\ Lett.\ B {\bf 115} (1982)
  111;
%
  R.~D'Auria, P.~Fre and T.~Regge,
  ``Consistent Supergravity In Six-Dimensions Without Action Invariance,''
  Phys.\ Lett.\ B {\bf 128} (1983) 44;
%
  Y.~Tanii,
  ``N=8 Supergravity In Six-Dimensions,''
  Phys.\ Lett.\ B {\bf 145} (1984) 197;
%
L.J. Romans, Nucl.\ Phys.\ {\bf B269} (1986) 691--711.

\bibitem{HiDSugra}
For a survey of many of the higher-dimensional supergravities see,
for example, {\it Supergravities in Diverse Dimensions} Vols. I \&
II, ed. by A. Salam and E. Sezgin, World Scientific 1989.

\bibitem{6DAnomalyCancellation}
S. Randjbar-Daemi, A. Salam, E. Sezgin and J. Strathdee, {\it
Phys. Lett.} {\bf B151} (1985) 351;
%
M.B. Green, J.H. Schwarz and P.C. West, {\it Nucl. Phys.} {\bf
B254} (1985) 327;
%
J. Erler, {\it J. Math. Phys.} {\bf 35} (1994) 1819
[hep-th/9304104].

\bibitem{Susha}
  Y.~Aghababaie, C.~P.~Burgess, S.~L.~Parameswaran and F.~Quevedo,
   ``Susy Breaking and Moduli Stabilization from Fluxes in Gauged
   6D
  Supergravity,''
  JHEP {\bf 0303} (2003) 032
  [hep-th/0212091].

\bibitem{ADD}
N. Arkani-Hamed, S. Dimopoulos and G. Dvali, Phys.\ Lett.\ B {\bf
429}, 263 (1998), [hep-ph/9803315]; Phys.\ Rev.\ D {\bf 59},
086004 (1999), [hep-ph/9807344];
%
I.~Antoniadis, N.~Arkani-Hamed, S.~Dimopoulos and G.~R.~Dvali,
``New dimensions at a millimeter to a Fermi and superstrings at a
TeV,'' Phys.\ Lett.\ B {\bf 436}, 257 (1998), [hep-ph/9804398].

\bibitem{WeakScaleGravity}
I.~Antoniadis, N.~Arkani-Hamed, S.~Dimopoulos and G.~Dvali, {\it
Phys.\ Lett.} {\bf B436} (1998) 257 [hep-ph/9804398];
%
P.~Horava and E.~Witten, {\it Nucl.\ Phys.} {\bf B475} (1996) 94
[hep-th/9603142]; {\it Nucl.\ Phys.} {\bf B460} (1996) 506
[hep-th/9510209];
%
E.~Witten, {\it Nucl.\ Phys.} {\bf B471} (1996) 135
[hep-th/9602070];
%
J. Lykken, {\it Phys.\ Rev.} {\bf D54} (1996) 3693
[hep-th/9603133];
%
I.~Antoniadis, {\it Phys.\ Lett.} {\bf B246} (1990) 377.
%
J.~J.~van der Bij, ``Large rescaling of the scalar condensate,
towards a Higgs gravity connection?,''
  [hep-ph/9908297].

\bibitem{6DSUSYBreaking}
D. Atwood, C.P. Burgess, E. Filotas, F. Leblond, D. London and I.
Maksymyk, Phys.\ Rev.\ D {\bf 63}, 025007 (2001),
[hep-ph/0007178];
%
J.~L.~Hewett and D.~Sadri, ``Supersymmetric extra dimensions:
Gravitino effects in selectron pair production,'' Phys.\ Rev.\ D
{\bf 69}, 015001 (2004);
%
  G.~Azuelos, P.~H.~Beauchemin and C.~P.~Burgess,
  ``Phenomenological constraints on extra-dimensional scalars,''
  J.\ Phys.\ G {\bf 31}, 1 (2005)
  [hep-ph/0401125];
  %
  C.~P.~Burgess, J.~Matias and F.~Quevedo,
  ``MSLED: A minimal supersymmetric large extra dimensions scenario,''
  Nucl.\ Phys.\ B {\bf 706} (2005) 71
  [hep-ph/0404135];
  %
   P.~H.~Beauchemin, G.~Azuelos and C.~P.~Burgess,
  ``Dimensionless coupling of bulk scalars at the LHC,''
  J.\ Phys.\ G {\bf 30}, N17 (2004)
  [hep-ph/0407196];
  %
  S.~L.~Parameswaran, S.~Randjbar-Daemi and A.~Salvio,
  ``Gauge fields, fermions and mass gaps in 6D brane worlds,''
  Nucl.\ Phys.\  B {\bf 767} (2007) 54
  [hep-th/0608074].

\bibitem{SLED1}
Y. Aghababaie, C.P. Burgess, S. Parameswaran and F. Quevedo,
Nucl.\ Phys.\ {\bf B680} (2004) 389--414, [hep-th/0304256];
%
  C.~P.~Burgess,
  ``Towards a natural theory of dark energy: Supersymmetric large extra
  dimensions,''
  AIP Conf.\ Proc.\  {\bf 743} (2005) 417
  [hep-th/0411140].

\bibitem{SLED2}
%
Y. Aghabababie, C.P. Burgess, J.M. Cline, H. Firouzjahi, S.
Parameswaran, F. Quevedo, G. Tasinato and I. Zavala, JHEP 0309
(2003) 037 [hep-th/0308064].

\bibitem{SLEDx}
%
C.P. Burgess, ``Supersymmetric Large Extra Dimensions and the
Cosmological Constant: An Update,'' {\it Ann. Phys.} {\bf 313}
(2004) 283-401 [hep-th/0402200];
%
  J.~Garriga and M.~Porrati,
  ``Football shaped extra dimensions and the absence of self-tuning,''
  JHEP {\bf 0408} (2004) 028
  [hep-th/0406158].

\bibitem{UVSensitivity}
 C.~P.~Burgess and D.~Hoover,
  ``UV sensitivity in supersymmetric large extra dimensions: The Ricci-flat
  case,''
  [hep-th/0504004];
  %
   D.~M.~Ghilencea, D.~Hoover, C.~P.~Burgess and F.~Quevedo,
  ``Casimir energies for 6D supergravities compactified on T(2)/Z(N) with
  Wilson lines,''
  JHEP {\bf 0509}, 050 (2005)
  [hep-th/0506164];
  %
  D.~Hoover and C.~P.~Burgess,
  ``Ultraviolet sensitivity in higher dimensions,''
  JHEP {\bf 0601}, 058 (2006)
  [hep-th/0507293];
  %
  E.~Elizalde, M.~Minamitsuji and W.~Naylor,
  ``Casimir effect in rugby-ball type flux compactifications,''
  Phys.\ Rev.\  D {\bf 75} (2007) 064032
  [hep-th/0702098].

\bibitem{SLEDpheno}
    J.~Matias and C.~P.~Burgess,
  ``MSLED, neutrino oscillations and the cosmological constant,''
  JHEP {\bf 0509} (2005) 052
  [hep-ph/0508156];
  %
   P.~Callin and C.~P.~Burgess,
  ``Deviations from Newton's law in supersymmetric large extra dimensions,''
  [hep-ph/0511216].

\bibitem{5DSelfTune}
N.~Arkani-Hamed, S.~Dimopoulos, N.~Kaloper and R.~Sundrum, ``A
small cosmological constant from a large extra dimension,'' Phys.\
Lett.\ B {\bf 480} (2000) 193, [hep-th/0001197];
%
S.~Kachru, M.~B.~Schulz and E.~Silverstein, ``Self-tuning flat
domain walls in 5d gravity and string theory,'' Phys.\ Rev.\ D
{\bf 62} (2000) 045021, [hep-th/0001206].

\bibitem{5DSelfTunex}
S.~Forste, Z.~Lalak, S.~Lavignac and H.~P.~Nilles, ``A comment on
self-tuning and vanishing cosmological constant in the  brane
world'', Phys.\ Lett.\ B {\bf 481} (2000) 360, hep-th/0002164;
JHEP {\bf 0009} (2000) 034, [hep-th/0006139];\\
%
C.~Csaki, J.~Erlich, C.~Grojean and T.J.~Hollowood, ``General
Properties of the Self-Tuning Domain Wall Approach to the
Cosmological Constant Problem,'' Nucl.\ Phys.\ {\bf B584} (2000)
359-386, [hep-th/0004133];\\
%
C.~Csaki, J.~Erlich and C.~Grojean, ``Gravitational Lorentz
Violations and Adjustment of the Cosmological Constant in
Asymmetrically Warped Spacetimes,'' Nucl.\ Phys.\ {\bf B604}
(2001) 312-342, [hep-th/0012143];\\
%
  C.~Grojean, F.~Quevedo, G.~Tasinato and I.~Zavala,
  ``Branes on charged dilatonic backgrounds: Self-tuning, Lorentz  violations and cosmology,''
  JHEP {\bf 0108} (2001) 005
  [arXiv:hep-th/0106120];\\
%
J.M. Cline and H. Firouzjahi, ``No-Go Theorem for Horizon-Shielded
Self-Tuning Singularities'', Phys.\ Rev.\ {\bf D65} (2002) 043501,
[hep-th/0107198].

\bibitem{6DNonSUSYSelfTune}
J.-W. Chen, M.A. Luty and E. Pont{\'o}n, JHEP 0009 (2000) 012,
[hep-th/0003067];
%
S.~M.~Carroll and M.~M.~Guica, ``Sidestepping the cosmological
constant with football-shaped extra dimensions,''
[hep-th/0302067];
%
I.~Navarro, ``Co-dimension two compactifications and the
cosmological constant  problem,'' JCAP {\bf 0309} (2003) 004
[hep-th/0302129].

\bibitem{6DNonSUSYSelfTunex}
I.~Navarro, ``Spheres, deficit angles and the cosmological
constant,'' Class.\ Quant.\ Grav.\  {\bf 20} (2003) 3603
[hep-th/0305014];
%
H.~P.~Nilles, A.~Papazoglou and G.~Tasinato, ``Selftuning and its
footprints,'' Nucl.\ Phys.\ B {\bf 677} (2004) 405
[hep-th/0309042];
%
  P.~Bostock, R.~Gregory, I.~Navarro and J.~Santiago,
  ``Einstein gravity on the Co-dimension 2 brane?,''
  Phys.\ Rev.\ Lett.\  {\bf 92}, 221601 (2004)
  [hep-th/0311074];
%
  J.~Vinet and J.~M.~Cline,
  ``Can Co-dimension-two branes solve the cosmological
  constant problem?,''
  Phys.\ Rev.\ D {\bf 70} (2004) 083514
  [hep-th/0406141];
  %
  M.~L.~Graesser, J.~E.~Kile and P.~Wang,
  ``Gravitational perturbations of a six dimensional
  self-tuning model,''
  Phys.\ Rev.\ D {\bf 70} (2004) 024008
  [hep-th/0403074];
%
  %
  G.~Kofinas,
  ``On braneworld cosmologies from six dimensions, and absence thereof,''
  [hep-th/0506035].

\bibitem{GGP}
  G.~W.~Gibbons, R.~Guven and C.~N.~Pope,
  ``3-branes and uniqueness of the Salam-Sezgin vacuum,''
  Phys.\ Lett.\ B {\bf 595}, 498 (2004)
  [hep-th/0307238].

\bibitem{GGPplus}
  C.~P.~Burgess, F.~Quevedo, G.~Tasinato and I.~Zavala,
  ``General axisymmetric solutions and self-tuning in 6D chiral gauged
  supergravity,''
  JHEP {\bf 0411}, 069 (2004)
  [hep-th/0408109].

\bibitem{RS}
  L.~Randall and R.~Sundrum,
  ``A large mass hierarchy from a small extra dimension,''
  Phys.\ Rev.\ Lett.\  {\bf 83} (1999) 3370
  [hep-ph/9905221];
  %
  ``An alternative to compactification,''
  Phys.\ Rev.\ Lett.\  {\bf 83} (1999) 4690
  [hep-th/9906064].

\bibitem{Gian1}
  M.~Peloso, L.~Sorbo and G.~Tasinato,
  ``Standard 4d gravity on a brane in six dimensional flux compactifications,''
  Phys.\ Rev.\  D {\bf 73} (2006) 104025
  [hep-th/0603026].

\bibitem{6DSmoothing}
  J.~Vinet and J.~M.~Cline,
  ``Codimension-two branes in six-dimensional supergravity and the
  cosmological constant problem,''
  Phys.\ Rev.\  D {\bf 71} (2005) 064011
  [hep-th/0501098];
%
  T.~Kobayashi and M.~Minamitsuji,
  ``Gravity on an extended brane in six-dimensional warped flux
  compactifications,''
  [hep-th/0703029];
%
  N.~Kaloper and D.~Kiley,
  ``Charting the landscape of modified gravity,''
  [hep-th/0703190].


\bibitem{Papazoglou}
  E.~Papantonopoulos, A.~Papazoglou and V.~Zamarias,
  ``Regularization of conical singularities in warped six-dimensional
  compactifications,''
  JHEP {\bf 0703} (2007) 002
  [arXiv:hep-th/0611311];
%
  H.~M.~Lee and A.~Papazoglou,
  ``Gravitino in six-dimensional warped supergravity,''
  arXiv:0705.1157 [hep-th].

\bibitem{OtherCap}
 B.~Himmetoglu and M.~Peloso,
  ``Isolated Minkowski vacua, and stability
  analysis for an extended brane in
  the rugby ball,''
  [hep-th/0612140].

\bibitem{GW}
  W.~D.~Goldberger and M.~B.~Wise,
  ``Modulus stabilization with bulk fields,''
  Phys.\ Rev.\ Lett.\  {\bf 83} (1999) 4922
  [hep-ph/9907447].

\bibitem{GandC}
S.~Weinberg, {\sl Gravitation and Cosmology}, Wiley, New York,
1972.

\bibitem{MTW}
C.W.~Misner, K.P.~Thorne and J.A.~Wheeler, {\sl Gravitation}, W.H.
Freeman and Company (1970).

\bibitem{HypersNonzero}
For solutions with nonzero hyperscalars see, however
%
   S.~Randjbar-Daemi and E.~Sezgin,
  ``Scalar potential and dyonic strings in 6d gauged supergravity,''
  Nucl.\ Phys.\ B {\bf 692} (2004) 346
  [hep-th/0402217];
%
 A.~Kehagias,
  ``A conical tear drop as a vacuum-energy drain for the solution of the
  cosmological constant problem,''
  Phys.\ Lett.\ B {\bf 600} (2004) 133
  [hep-th/0406025];
%
  S.~Randjbar-Daemi and V.~A.~Rubakov,
  ``4d-flat compactifications with brane vorticities,''
  JHEP {\bf 0410}, 054 (2004)
  [hep-th/0407176];
%
  H.~M.~Lee and A.~Papazoglou,
  ``Brane solutions of a spherical sigma model in six dimensions,''
  Nucl.\ Phys.\ B {\bf 705} (2005) 152
  [hep-th/0407208];
%
 V.~P.~Nair and S.~Randjbar-Daemi,
  ``Nonsingular 4d-flat branes in six-dimensional supergravities,''
  JHEP {\bf 0503} (2005) 049
  [hep-th/0408063];
%
   S.~L.~Parameswaran, G.~Tasinato and I.~Zavala,
  ``The 6D SuperSwirl,''
  [hep-th/0509061];
%
 H.~M.~Lee and C.~Ludeling,
  ``The general warped solution with conical branes in six-dimensional
  supergravity,''
  [hep-th/0510026].

\bibitem{6DdSSUSY}
  A.~Tolley, C.P.~Burgess, D.~Hoover and Y.~Aghababaie,
  ``Bulk Singularities and the Effective Cosmological Constant
   for Higher Co-dimension Branes,''
   JHEP {\bf 0603} (2006) 091 [hep-th/0512218].

\bibitem{Linearized}
    J.M.~Cline, J.~Descheneau, M.~Giovannini and J.~Vinet, JHEP 0306
    (2003) 048 [hep-th/0304147];
%
  H.~M.~Lee and A.~Papazoglou,
  [hep-th/0602208];

\bibitem{KickRB}
    C.~P.~Burgess, C.~de Rham, D.~Hoover, D.~Mason and A.~J.~Tolley,
  ``Kicking the rugby ball: Perturbations of 6D gauged chiral supergravity,''
  JCAP {\bf 0702} (2007) 009
  [hep-th/0610078].

\bibitem{Scaling}
  A.~J.~Tolley, C.~P.~Burgess, C.~de Rham and D.~Hoover,
  ``Scaling solutions to 6D gauged chiral supergravity,''
  New J.\ Phys.\  {\bf 8} (2006) 324
  [hep-th/0608083].

\bibitem{GibbonsHawking}
G.W.~Gibbons and S.W.~Hawking, Phys.\ Rev.\ {\bf D15} (1977) 2752.

\bibitem{NavSant}
  I.~Navarro and J.~Santiago,
  ``Gravity on Co-dimension 2 brane worlds,''
  JHEP {\bf 0502}, 007 (2005)
  [hep-th/0411250].

\bibitem{GP}
G.~W.~Gibbons and C.~N.~Pope,
   ``Consistent S**2 Pauli reduction of six-dimensional
   chiral gauged
  Einstein-Maxwell supergravity,''
  Nucl.\ Phys.\ B {\bf 697} (2004) 225
  [hep-th/0307052].

  \bibitem{Israel}
  K.~Lanczos, Phys.\ Z.\ {\bf 23} (1922) 239--543; Ann.\ Phys.\
  {\bf 74} (1924) 518--540;
  %
  C.W.~Misner and D.H.~Sharp, Phys.\ Rev.\ {\bf 136} (1964)
  571--576;
  %
  W.~Israel, Nuov.\ Cim.\ {\bf 44B} (1966) 1--14; {\it errata}
  Nuov.\ Cim.\ {\bf 48B} 463.


\bibitem{EoWRG}
  J.~de Boer, E.~P.~Verlinde and H.~L.~Verlinde,
  ``On the holographic renormalization group,''
  JHEP {\bf 0008} (2000) 003
  [hep-th/9912012].

\bibitem{Inprogress}
C.P.~Burgess, D.~Hoover and G.~Tasinato (in preparation).

\end{thebibliography}
\end{document}